\documentclass[sigconf,nonacm]{acmart}

\usepackage{amsmath}

\usepackage{multirow,multicol,booktabs,tabulary,tabu,longtable,array,varwidth}
\usepackage{placeins, lipsum}
\usepackage[flushleft]{threeparttable}
\usepackage{tablefootnote}
\usepackage{flushend}

\usepackage{graphicx}
\usepackage{pgfplots}
\usepackage{colortbl}
\usepackage{siunitx}

\usepackage{enumerate}

\usepackage[inline]{enumitem}
\setlist{noitemsep,nolistsep,leftmargin=*}

\usepackage{algorithm}
\usepackage{algpseudocode}

\usepackage{xspace}
\usepackage{mdframed}

\usepackage[labelfont=bf]{caption}
\usepackage{subcaption}
\usepackage[normalem]{ulem}

\usepackage[capitalize,nameinlink]{cleveref}

\newcommand{\jheading}[1]{\vspace{0.05in}\noindent\textbf{#1.}}

\newcommand{\myparagraph}[1]{\vspace{\smallskipamount}\noindent\textbf{#1.\xspace}}

\newcommand{\myparagraphemph}[1]{\vspace{\smallskipamount}\noindent\emph{#1.\xspace}}
\newcommand{\eg}{\emph{e.g.}\xspace}

\newcommand{\ie}{\emph{i.e.}\xspace}
\newcommand{\etal}{\emph{et al.}\xspace}
\newcommand*{\rom}[1]{\uppercase\expandafter{\romannumeral #1\relax}}

\crefname{insight}{Insight}{Insight}
\newcounter{insight}
\newcommand{\boxinsight}[1]{%
  \refstepcounter{insight}%
  \vspace{\smallskipamount}%
  \noindent \simplebox{
    \textbf{\uline{Insight~\theinsight:}} 
    \textit{#1}
  }
}

\usepackage{ifthen}
\newboolean{publicversion}
\setboolean{publicversion}{false}

\ifthenelse{\boolean{publicversion}}{
    \newcommand{\grumbler}[3]{}
    \newcommand{\esha}[1]{}
    \newcommand{\inigo}[1]{}
    \newcommand{\haoran}[1]{}
    \newcommand{\alind}[1]{}
    \newcommand{\anish}[1]{}
    \newcommand{\rr}[1]{}
    \newcommand{\hq}[1]{}
    \newcommand{\todo}[1]{}
}
{
    \newcommand{\grumbler}[3]{\xspace\textcolor{#3}{\bf #1: #2}}
    \newcommand{\esha}[1]{\grumbler{Esha}{#1}{brown}}
    \newcommand{\inigo}[1]{\grumbler{Inigo}{#1}{violet}}
    \newcommand{\haoran}[1]{\grumbler{Haoran}{#1}{blue}}
    
    \newcommand{\alind}[1]{\grumbler{Alind}{#1}{ForestGreen}}
    \newcommand{\anish}[1]{\grumbler{Anish}{#1}{teal}}
    \newcommand{\rr}[1]{\grumbler{Ram}{#1}{purple}}
    \newcommand{\hq}[1]{\textcolor{blue}{hqiu: #1}}
    \newcommand{\todo}[1]{\textcolor{blue}{TODO: #1}}
}

\hypersetup{
    unicode=true,
    bookmarksnumbered=true,
    colorlinks=true,
    linkcolor={red!70!black},
    citecolor={red!70!black},
    urlcolor={blue!70!black},
    pdfborder={0 0 0}
}

\usepackage[most]{tcolorbox}
\tcbset{textmarker/.style={%
        enhanced,
        parbox=false,boxrule=0mm,boxsep=0mm,arc=1.5mm,
        outer arc=1.5mm,left=2mm,right=2mm,top=4pt,bottom=3pt,
        toptitle=1mm,bottomtitle=1mm,oversize}}

\newtcolorbox{simplenoteBox}{colback=white, colframe=black, boxrule=0.2mm, arc=0mm, auto outer arc, boxsep=0mm, left=2mm, right=2mm, top=1mm, bottom=1mm} 
\newtcolorbox{noteBox}{textmarker,
    colback=gray!8!white}

\newcommand{\simplebox}[1]{\begin{simplenoteBox} #1 \end{simplenoteBox}}

\definecolor{mygray}{gray}{0.95}
\newmdenv[
  backgroundcolor=mygray,
  linecolor=black,
  linewidth=1pt,
  roundcorner=3pt,
  nobreak=true,
]{keyfinding}

\newcommand{\lmm}{\textsc{LMM}\xspace}
\newcommand{\lmms}{\textsc{LMM}s\xspace}
\newcommand{\sysname}{\textsc{ModServe}\xspace}

\newcommand{\provider}{{Azure}\xspace}

\begin{document}

\title{\sysname: Modality- and Stage-Aware Resource Disaggregation for Scalable Multimodal Model Serving}

\author[]{Haoran Qiu}
\affiliation{Microsoft Azure Research
\country{USA}}
\email{haoran.qiu@microsoft.com}

\author[]{Anish Biswas}
\affiliation{Microsoft Research India
\country{India}}
\email{t-anibiswas@microsoft.com}

\author[]{Zihan Zhao}
\affiliation{University of Virginia
\country{USA}}
\email{rxy6cc@virginia.edu}

\author[]{Jayashree Mohan}
\affiliation{Microsoft Research India
\country{India}}
\email{jamohan@microsoft.com}

\author[]{Alind Khare}
\affiliation{Microsoft M365 Research
\country{India}}
\email{alindkhare@microsoft.com}

\author[]{Esha Choukse}
\affiliation{Microsoft Azure Research
\country{USA}}
\email{esha.choukse@microsoft.com}

\author[]{Íñigo Goiri}
\affiliation{Microsoft Azure Research
\country{USA}}
\email{inigog@microsoft.com}

\author[]{Zeyu Zhang}
\affiliation{University of Virginia
\country{USA}}
\email{qxc4fh@virginia.edu}

\author[]{Haiying Shen}
\affiliation{University of Virginia
\country{USA}}
\email{hs6ms@virginia.edu}

\author[USA]{Chetan Bansal}
\affiliation{Microsoft M365 Research
\country{USA}}
\email{chetanb@microsoft.com}

\author[]{Ramachandran Ramjee}
\affiliation{Microsoft Research India
\country{India}}
\email{ramjee@microsoft.com}

\author[]{Rodrigo Fonseca}
\affiliation{Microsoft Azure Research
\country{USA}}
\email{fonseca.rodrigo@microsoft.com}

\renewcommand{\shortauthors}{Haoran Qiu et al.}

\begin{abstract}
Large multimodal models (\lmms{}) demonstrate impressive capabilities in understanding images, videos, and audio beyond text.
However, efficiently serving \lmms{} in production environments poses significant challenges due to their complex model architectures and heterogeneous characteristics across their multi-stage inference pipelines and modalities.

We present the first comprehensive systems analysis of two prominent \lmm architectures, decoder-only and cross-attention, across six representative open-source models, revealing key systems design implications.
We also present an in-depth analysis of production \lmm inference traces, uncovering unique multimodal workload characteristics, including variable, heavy-tailed request distributions and bursty traffic patterns.

Based on these insights, we propose \sysname{}, a modular \lmm{} serving system that decouples model stages for independent optimization and adaptive scaling. \sysname{} dynamically reconfigures stages and handles bursty traffic with modality-aware scheduling and autoscaling to meet tail latency SLOs while minimizing costs.
\sysname{} achieves 3.3--5.5$\times$ higher throughput (leading to 25--41.3\% cost saving) while meeting SLOs on a 128-GPU cluster with production multimodal traces.
\end{abstract}

\maketitle 

\section{Introduction}

\noindent
The rapid advancement in generative AI has led to the development of large multimodal models (\lmms{}) capable of processing inputs across various modalities such as text, image, video, and audio.
These models have demonstrated remarkable capabilities in tasks like image captioning~\cite{chen2022visualgpt,mokady2021clipcap,hu2023promptcap}, visual question answering~\cite{schwenk2022okvqa,shao2023prompting}, multimodal dialogue systems~\cite{li2024llava,chen2024internvl,team2024gemini}, and voice assistant~\cite{he2025deployment}.
This has led to a rapid adoption of \lmms{} in production services, including online user-facing applications where latency service-level objectives (SLOs) are critical.

Unlike traditional large language models (LLMs) that process purely textual inputs using a single component, a decoder-based transformer architecture~\cite{transformer}, \lmms{} handle fundamentally different types of inputs, each requiring distinct processing approaches.
This heterogeneity introduces unique serving complexities that demand novel analysis and serving strategies.
For \textit{Image-Text-to-Text} models~\cite{ittt}, the inference pipeline consists of multiple specialized stages:
image preprocessing to transform raw images into tensor representations, image encoding to convert these tensors into image tokens, and a language model backend that combines text prompts with image tokens to generate text outputs.
Currently, these stages are typically served as a monolithic system~\cite{vllm,hf,deepspeed}, where all components are integrated within a single serving instance and scaled together as a unified entity.
While recent LLM serving systems adopt prefill-decode (PD) disaggregation~\cite{patel2024splitwise,zhong2024distserve} to reduce the performance interference at the LLM backend, these optimizations remain text-centric and overlook the upstream stages of multimodal preprocessing and encoding, which are still tightly coupled within a monolithic serving instance.

\begin{figure}[!t]
    \raggedright
    \includegraphics[width=0.97\linewidth]{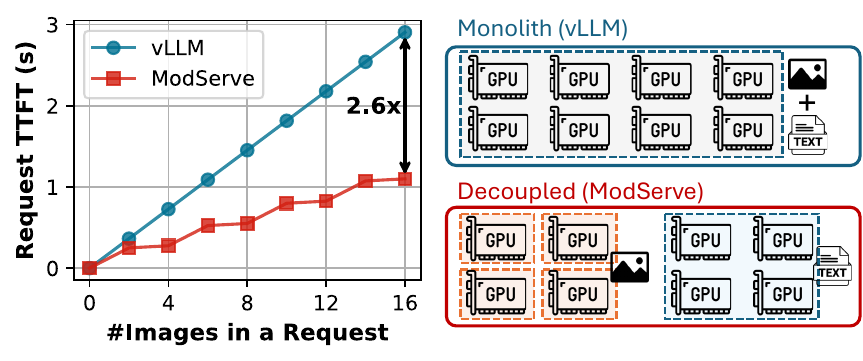}
    \caption{Impact of image/video workload on \lmm{} inference TTFT for state-of-the-art implementation of Llama3.2-11B on vLLM vs. \sysname{} with an 8-A100 GPU server.
    The ``Monolith'' setup deploys the full model using 8 GPUs while the ``Decoupled'' setup deploys the LLM backend on 4 GPUs and four image encoders on the other 4 GPUs.
    \vspace{-15pt}
    }
    \label{fig:monolith-vs-decoupled}
\end{figure}

This lack of modality-aware decoupling limits the efficiency and scalability of existing monolithic inference serving systems when serving workloads beyond text. These systems struggle to meet time-to-first-token (TTFT) SLOs because multimodal input preprocessing and encoding are in the critical path.
\Cref{fig:monolith-vs-decoupled} illustrates the challenges faced by a monolithic deployment in scaling as the number of images per request increases (a common scenario in multi-image or video workloads), resulting in sharp TTFT degradation.
As a result, image-heavy requests can result in head-of-line (HoL) blocking, reducing system responsiveness and causing resource overprovisioning.

\jheading{Our Work}
In this paper, we present the first comprehensive systems analysis of two leading \lmm{} architectures:
cross-attention (\emph{CroAttn}) and decoder-only (\emph{DecOnly}), on both open-source \lmms{} and novel production \lmm{} inference traces in \provider datacenters. We analyze their multi-stage inference pipelines, performance-resource tradeoffs, and production workload patterns, including variable request rates, diverse multimodal inputs, and bursty traffic.
To advance research in this area, we release the \textbf{\textit{first open-source multimodal trace}} from our production clusters, enabling the community to study real-world deployment patterns.

Our analysis identifies three key insights for optimizing \lmm{} inference.
First, different \lmm{} inference stages exhibit diverse performance characteristics and varying sensitivity to resource and model configurations (\eg{}, batching and model sharding), necessitating \emph{{decoupled execution}}.
Second, image encoding is a major bottleneck for TTFT, requiring efficient \emph{{encoder parallelization}} to reduce both latency and HoL blocking.
Finally, production multimodal traffic exhibits distinct bursty patterns driven by increased images per request, highlighting the need for \emph{{modality-aware routing}} strategies to manage bursts and mitigate tail latency spikes.

Based on these insights, we propose \sysname{}, a novel \textit{\textbf{modular architecture}} for scalable and resource-efficient \lmm{} serving which directly addresses the challenges identified in our analysis.
Take Image-Text-to-Text tasks as an example, \sysname{} separates image- and text-specific inference stages into distinct instances for decoupled execution.
In \sysname{}, \emph{Image Instances} handle image preprocessing and encoding, while \emph{Text Instances} manage LLM prefill and decoding (\Cref{fig:monolith-vs-decoupled}).
Text-only requests are served by \emph{Text Instances}, whereas image-text requests go through \emph{Image Instances} where images are converted to tokens before being forwarded to \emph{Text Instances} for text generation.

\sysname{}'s modular architecture unlocks stage-specific optimizations. \sysname{} manages \emph{Image} and \emph{Text Instances} independently with stage-aware autoscaling, model sharding, and batching.
By autoscaling the stages separately, it minimizes resource overprovisioning.
For example, during image-driven bursts observed in production traffic, \emph{Image Instances} can scale out independently, making \sysname{} more resource-efficient than monolithic inference systems.
To navigate the image encoding bottleneck, \sysname{} parallelizes encoding of a single request across multiple \emph{Image Instances} (\Cref{fig:monolith-vs-decoupled}), leveraging our finding that the images within a request do not attend to each other during encoding, and hence the requests can be parallelized at the image level.

Further, to manage image-driven bursts, \sysname{} implements modality-aware request routing for \emph{Image and Text Instances}.
For example, images from image-text requests are routed to \emph{Image Instances} with the fewest pending image tokens to encode, reducing HoL blocking and tail latency spikes.

We implement \sysname{} on top of a high-performance inference system, vLLM~\cite{vllm}, and demonstrate the effectiveness of \sysname{} through extensive evaluations on a 16-server (128 GPUs) cluster running production \lmm{} inference traces from \provider.
Compared to state-of-the-art baselines, \sysname{} achieves \textbf{\textit{3.3--5.5$\times$ higher throughput}} under static allocation and \textbf{\textit{reduces \lmm{} serving cost by 25--41.3\%}} while meeting the P99 TTFT SLOs.

While existing techniques like PD disaggregation have shown promise for optimizing LLM inference by separating prefill and decode phases~\cite{zhong2024distserve, patel2024splitwise}, they fall short for multimodal workloads. The unique nature of bursty and heterogeneous multimodal traffic observed in production makes it nontrivial to extend PD disaggregation studied in the context of text-centric LLM inference; modality-specific optimizations such as parallelizing encoding, modality-aware routing, and stage autoscaling are crucial to meet SLOs while maintaining resource efficiency. Additionally, the choice of selecting PD disaggregation or colocation for the LLM backend depends on the workload conditions and optimization targets~\cite{zhong2024distserve}. We show in ~\cref{sec:eval} that \sysname is composable with both colocated LLM backends (mixed PD batching) as well as PD disaggregation for the text nodes, and improves serving latency with the proposed modality-aware techniques in both LLM backend configurations.

We focus on Image-Text-to-Text and Video-Text-to-Text (where videos are processed as image frame sequences~\cite{llava-ov}), but our insights extend to other multimodal scenarios, such as Audio-Text-to-Text tasks~\cite{attt}, which share similar model architectures and inference stages with the models we study.

\myparagraph{Summary}
This paper makes the following contributions:
\begin{itemize}[leftmargin=*]
\item The first open-source dataset containing large-scale production \lmm{} inference traces from \provider \cite{oss}.
\item A comprehensive system characterization on \lmm{} serving, examining performance profiles and resource utilization patterns across diverse multimodal workloads in both open-source \lmm deployments and production environments.
\item Design and implementation of \sysname{}, a modular architecture for scalable and resource-efficient \lmm{} serving.
\item A thorough evaluation of \sysname{} in a 128-GPU cluster using large-scale production traces.
\end{itemize}
\section{Large Multimodal Models Background}
\label{sec:background:lmm}

\begin{figure}[!t]
    \centering
    \includegraphics[width=0.86\linewidth]{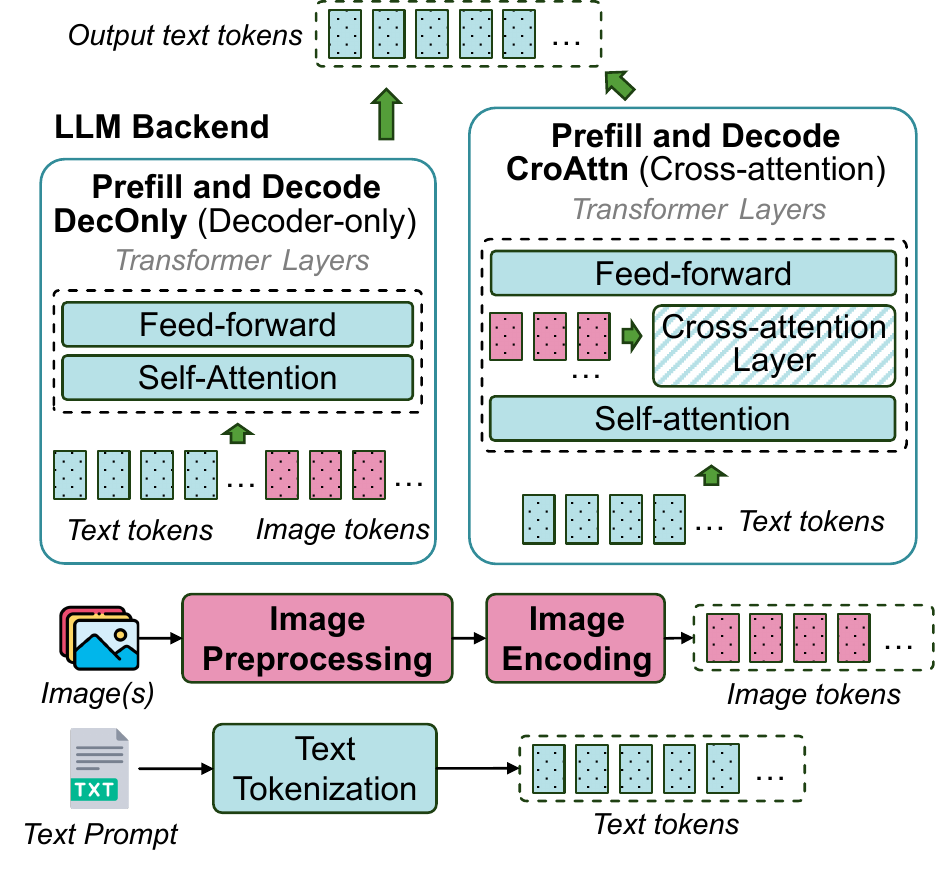}
    \caption{Model architecture for decoder-only and cross-attention-based \lmms{} in Image-Text-to-Text tasks~\cite{ittt}.}
    \label{fig:lmm-background}
\end{figure}

\lmms{} extend text-centric LLMs by integrating multimodal understanding capabilities for tasks like visual question answering~\cite{shao2023prompting} and computer-using agents~\cite{cua2025,niu2024screenagent}.
\Cref{fig:lmm-background} shows the typical pipeline of \lmm{} inference in visual understanding tasks~\cite{ittt}, which consists of three key stages: (1) \emph{image preprocessing}, where raw images are transformed into uniform-sized tiles; (2) \emph{image encoding}, where an encoder extracts visual features and produces a sequence of image tokens; and (3) \emph{text generation}, where an LLM backend processes the image and text tokens to generate output text tokens.
There are two dominant \lmm{} architectures that differ in how the LLM backend handles image tokens and text tokens:
(1) \emph{decoder-only} (DecOnly), used in models like DeepSeek's Janus~\cite{janus}, LLaVA-OneVision~\cite{llava-ov}, InternVL~\cite{chen2024internvl}, and NVLM-D~\cite{nvlm}; and (2) \emph{cross-attention-based} (CroAttn), found in Llama-3.2 Vision~\cite{llama3}, NVLM-X~\cite{nvlm}, and Flamingo~\cite{alayrac2022flamingo}.
In this work, we analyze six open-source \lmms{} (listed in \Cref{table:model-config}) across these architectures, varying image encoder sizes (400M–6B) and LLM scales (7B–72B).

\begin{table*}[tb!]
\caption{Model configurations for six representative open-source \lmms{} with an example input image of $896 \times 896$ pixels.}
\centering
\setlength{\tabcolsep}{2pt}
\setlength{\aboverulesep}{2pt}
\setlength{\belowrulesep}{2pt}
\setlength{\extrarowheight}{0pt}
\resizebox{0.97\linewidth}{!}{%
\begin{tabular}{l|l|c|c|c|c|c|c|c}
\toprule
\multirow{2}{*}{\textbf{\lmm Model Name}} &
\multirow{2}{*}{\textbf{Abbreviation}} &
\multirow{2}{*}{\textbf{Architecture}} &
\multirow{2}{*}{\textbf{Tile Size}} &
\textbf{Image Encoder} &
\textbf{Total Image Token Size} &
\textbf{LLM Backend} &
\textbf{Tensor} &
\textbf{Avgerage Accuracy}
\\
& & & &
{(\#Params)} &
\textbf{($\#Tiles\times{}\#TokensPerTile$)} &
{(\#Params)} &
\textbf{Parallelism} &
{(Open VLM Benchmark~\cite{vlm-leaderboard})}\\
\midrule
Llama 3.2 Vision 11B ~\cite{llama-3.2-11b-instruct} & Llama3.2-11B & Cross-attention & 560$\times$560 & ViT-H/14 (630M)  & 4 $\times$ 1601 $\times$ 1 = 6404 & Llama 3.1 (8B) & TP-4 & 
\begin{tikzpicture}
    \fill[mygray] (0,2.5) rectangle (5,3);
    \fill[red!20] (0,2.5) rectangle (2.865,3);
    \node[right] at (2.9,2.75) {57.8\%};
\end{tikzpicture}\\
Llama 3.2 Vision 90B~\cite{llama-3.2-90b-instruct} & Llama3.2-90B & Cross-attention & 560$\times$560 & ViT-H/14 (630M) & 4 $\times$ 1601 $\times$ 1 = 6404 & Llama 3.1 (70B) & TP-8 & 
\begin{tikzpicture}
    \fill[mygray] (0,2.5) rectangle (5,3);
    \fill[orange!20] (0,2.5) rectangle (3.17,3);
    \node[right] at (3.18,2.75) {63.4\%};
\end{tikzpicture} \\
LLaVA-OneVision 7B~\cite{llava-7b} & LLaVA-OV-7B & Decoder-only & 384$\times$384 & SigLIP (400M)  & 10 $\times$ 729 $\times$ 1 = 7290 & Qwen2 (7B) & TP-4 & \begin{tikzpicture}
    \fill[mygray] (0,2.5) rectangle (5,3);
    \fill[yellow!30] (0,2.5) rectangle (3.005,3);
    \node[right] at (3.11,2.75) {60.1\%};
\end{tikzpicture} \\
LLaVA-OneVision 72B~\cite{llava-72b} & LLaVA-OV-72B & Decoder-only & 384$\times$384 & SigLIP (400M)  & 10 $\times$ 729 $\times$ 1 = 7290 & Qwen2 (72B) & TP-8 & \begin{tikzpicture}
    \fill[mygray] (0,2.5) rectangle (5,3);
    \fill[orange!20] (0,2.5) rectangle (3.4,3);
    \node[right] at (3.5,2.75) {68\%};
\end{tikzpicture}\\
InternVL-2.5 26B~\cite{internvl-26b} & InternVL-26B & Decoder-only & 448$\times$448 & InternViT (6B)& $5 \times 256$ = 1280 & InternLM (20B) & TP-8 & \begin{tikzpicture}
    \fill[mygray] (0,2.5) rectangle (5,3);
    \fill[green!20] (0,2.5) rectangle (3.58,3);
    \node[right] at (3.7,2.75) {71.6\%};
\end{tikzpicture} \\
NVLM-D 72B~\cite{nvlm-72b} & NVLM-D-72B & Decoder-only & 448$\times$448 & InternViT (6B) & $5 \times 256$ = 1280 & Qwen2-Instruct (72B) & TP-8 & \begin{tikzpicture}
    \fill[mygray] (0,2.5) rectangle (5,3);
    \fill[orange!20] (0,2.5) rectangle (3.38,3);
    \node[right] at (3.4,2.75) {67.6\%};
\end{tikzpicture}\\
\bottomrule
\end{tabular}}
\label{table:model-config}
\end{table*}


\myparagraph{Image Preprocessing}
\lmms{} typically follow three preprocessing steps on CPU:
(1) resize, rescale, pad, and normalize the raw image, (2) segment it into tiles~\cite{llama3,nvlm,chen2024internvl} or patches~\cite{llava-ov}, and (3) apply tile/patch-level transformations and sampling.
The number of tiles varies, with higher image dimensions resulting in more tiles, which ultimately increases the number of image tokens.
For example, an image with 896\texttimes{}896 pixels generates 4, 5, or 10 tiles of different sizes after preprocessing for six open-source \lmms{} (\Cref{table:model-config}).

\begin{figure}[!t]
    \centering
    \includegraphics[width=0.95\linewidth]{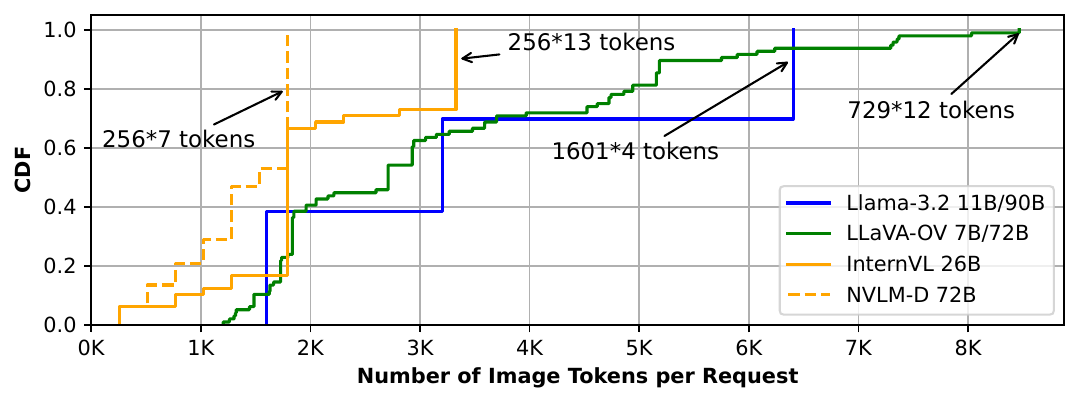}
    \caption{Distribution of image token count (per request) for open-source \lmms{} on ShareGPT-4o dataset~\cite{chen2024far}.
    Different \lmms{} (\eg{}, LLaVA-OV 7B and 72B) can share the same image encoder so the number of image tokens is the same.}
    \label{fig:img-token-distribution-open}
\end{figure}

\myparagraph{Image Encoding}
The image encoder takes processed image tiles as input and produces image tokens that are then passed to the language model backend.
Today's image encoders predominantly use the vision transformer architecture~\cite{alexey2020image} to extract visual features from images.
\Cref{table:model-config} shows that different \lmms{} use different encoders~\cite{llava-ov,alexey2020image,zhai_sigmoid_2023,chen2024internvl}, leading to variations in the number of image tokens when running image encoders on the same ShareGPT-4o dataset~\cite{chen2024far} (\Cref{fig:img-token-distribution-open}).
This is due to differences in the number of tiles and image tokens generated per tile by each encoder.

\myparagraph{Text Generation}
Image and text prompt tokens are combined and passed through LLM prefill and decode to generate output tokens, typically using one of two architectures:

\myparagraphemph{Decoder-Only (DecOnly) \lmms{}}
An unmodified LLM backend is reused in DecOnly \lmms{} (\eg{}, LLaVA-OV reuses Qwen2 LLM~\cite{llava-ov}), processing text and image tokens uniformly (shown as the ``DecOnly'' box in \Cref{fig:lmm-background}).
This works by attaching a connector~\cite{zhu2025connector} or modality-alignment module (\eg{}, MLP) that maps the image encoder output into the LLM’s token space.
While valued for their simplicity and unified modality handling, DecOnly models often require long sequences for high-resolution images, resulting in computational inefficiencies during inference.

\myparagraphemph{Cross-Attention (CroAttn) \lmms{}}
Unlike DecOnly \lmms{}, which leave the LLM backend unchanged, CroAttn-based models (\eg{}, Llama-3.2 Vision) integrate \emph{cross-attention layers} to process image tokens, treating visual inputs like a ``foreign language'' in the LLM backend.
While more complex to train, they improve inference efficiency by avoiding full image token unrolling in the LLM decoder, making them ideal for high-resolution inputs.
Self-attention operates on text tokens, while the cross-attention layer attends to both text and image tokens (``CroAttn'' box in \Cref{fig:lmm-background}).

\myparagraph{SLO Metrics for \lmm{} Inference}
Production \lmm{} serving systems need to satisfy SLOs defined on tail latency (\eg{}, P99) for worst-case performance.
These SLO metrics include \emph{Time to First Token (TTFT)} and \emph{Time Between Tokens (TBT)}.
TTFT measures the latency from query (text/images) to the first response token, critical for interactive applications.
In contrast to text-centric LLM serving, \lmm{} TTFT includes the following stages of \lmms{} inference pipeline:
(1) image preprocessing,
(2) encoding, and
(3) language model prefill time.
TBT captures the delay between consecutive token generations during decoding, affecting output fluency.
As multimodal preprocessing and encoding primarily influence TTFT, in this work, we focus on TTFT while leveraging state-of-the-art techniques~\cite{sarathi-serve,patel2024splitwise} to optimize TBT.
An ideal \lmm{}-serving system should meet TTFT/TBT SLOs while maximizing request \emph{throughput} (\ie{}, goodput) and compute \emph{utilization} (GPU cost).


\myparagraph{\lmm{} Deployments Today}
State-of-the-art serving frameworks~\cite{vllm,hf,deepspeed} deploy \lmms{} as \emph{monolithic} systems to meet latency SLOs.
In this setup, all inference components (\ie{}, image preprocessor, image encoder, and LLM backend) are co-located on the same hardware server as a single unit.
While PD disaggregation can be applied within this monolithic setup to decouple prefill from decode phases and reduce interference during text generation, it still couples multimodal components with the prefill instances~\cite{vllm}, preventing independent optimization of each component based on their distinct resource requirements and performance characteristics.
These tightly coupled components share uniform batching and model parallelism strategies across the pipeline.
\Cref{table:model-config} details the default model parallelism for our open-source \lmms{}.
While this monolithic design is straightforward to implement and common in open-source \lmm{} serving, it limits flexibility and suffers from sharp TTFT degradation under image-heavy workloads (\Cref{fig:monolith-vs-decoupled}).
\section{Motivation and \lmm{} Characterization}
\label{sec:characterization}

To further understand the limitations of monolithic deployments and explore unique characteristics that distinguish \lmm{} serving from text-centric LLM serving, we characterize open-source \lmms{} in the \emph{Image-Text-to-Text} category~\cite{ittt}.
We evaluate the performance and resource characteristics of heterogeneous inference stages under varying image inputs and model configurations (\Cref{sec:characterization:open-source}).
Moreover, to understand multimodal traffic patterns at scale, we analyze sample production traces from one production \lmm{} inference cluster at \provider (\Cref{sec:characterization:production}).

\begin{figure}[!t]
    \centering
    \includegraphics[width=\linewidth]{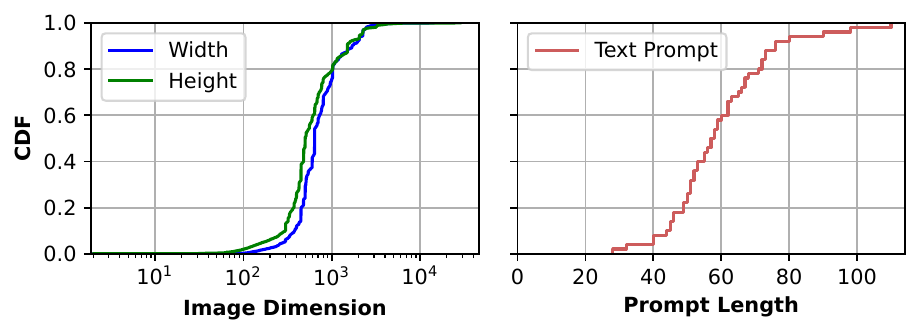}
    \caption{Image dimension distribution and text prompt length distribution of ShareGPT-4o Image dataset~\cite{chen2024far}.
    }
    \label{fig:sharegpt}
\end{figure}

\begin{figure*}[!t]
    \centering
    \includegraphics[width=\linewidth]{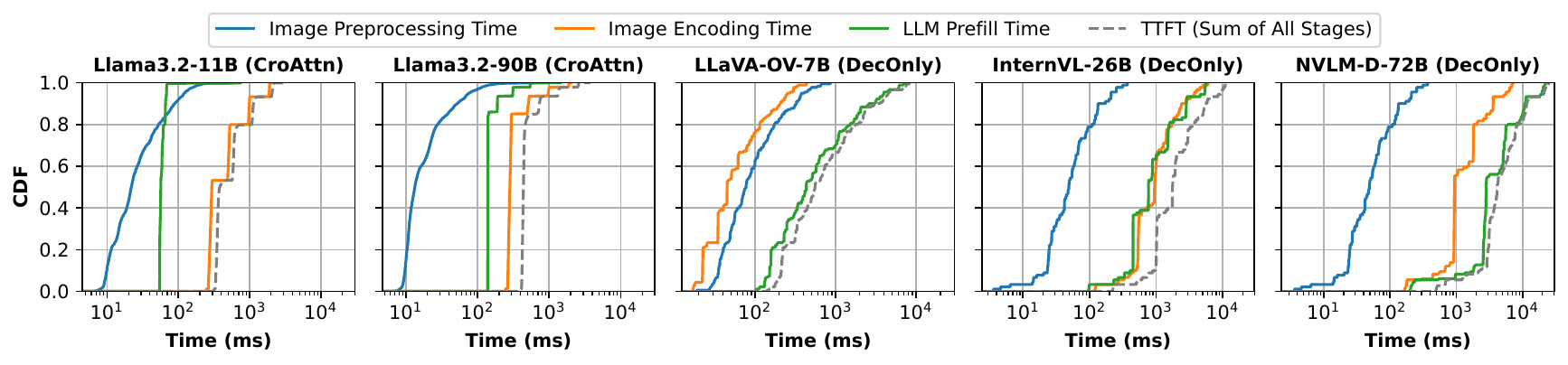}
    \caption{Per-stage request latency breakdown analysis across representative open-source \lmms{} deployed using default tensor parallelism (TP) as described in \Cref{table:model-config}.
    TTFT (dashed line) is the sum of the latency from each inference stage.
    }
    \vspace{-10pt}
    \label{fig:latency_breakdown}
\end{figure*}

\myparagraph{Characterization Setup}
The following is our setup:

\myparagraphemph{Models}
We use six representative open-source \lmms{} across two different architectures (DecOnly and CroAttn) as listed in \Cref{table:model-config}.
We deploy the models on vLLM~\cite{vllm} in BF16 with the default, stable PD colocated mode because the experimental PD disaggregated mode of vLLM has limited support for multimodality, and no support for cross-attention models. However, we later show in \cref{sec:eval:pd-disaggregation} that PD disaggregation is an optimization that is trivially composable with our design in \sysname.

\myparagraphemph{Dataset}
We use the open-source ShareGPT-4o dataset~\cite{chen2024far}, which includes 50K images of varying resolutions and text prompts from multimodal GPT-4o as shown in \Cref{fig:sharegpt}.

\myparagraphemph{Hardware}
Our setup features a DGX-A100 server with 8 NVIDIA A100 GPUs (80GB each) connected via NVLINK~\cite{a100azure}.
It has 96 AMD Epyc™ 7V12 CPU cores and 1900 GiB DRAM.

\subsection{Characterization on Open-Source \lmms}
\label{sec:characterization:open-source}
We characterize open-source \lmms{} to understand how different inference stages impact performance and resource efficiency.
Additionally, we compare DecOnly and CroAttn models to highlight the need for model-specific optimization.

\begin{figure*}[!t]
  \centering
  \begin{subfigure}[b]{0.22\textwidth}
    \centering
    \includegraphics[width=1\textwidth]{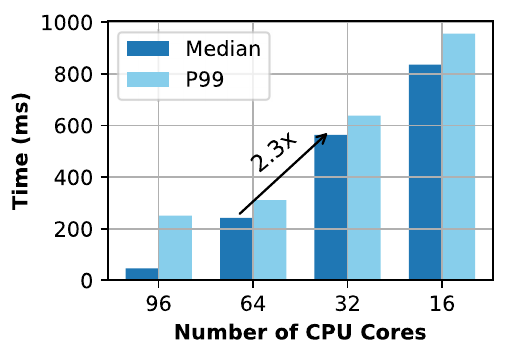}
    \caption{Image preprocessing time varying CPU allocation.}
    \label{fig:vary-preprocessor-core}
  \end{subfigure}%
  \hfill
  \begin{subfigure}[b]{0.21\textwidth}
    \centering
    \includegraphics[width=\textwidth]{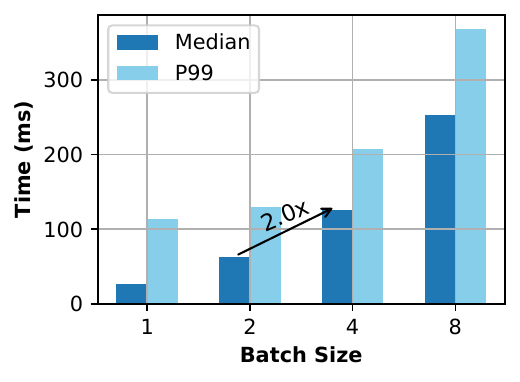}
    \caption{Image preprocessing time varying batch size.}
    \label{fig:vary-img-preprocessor-batch}
  \end{subfigure}%
  \hfill
  \begin{subfigure}[b]{0.21\textwidth}
    \centering
    \includegraphics[width=\textwidth]{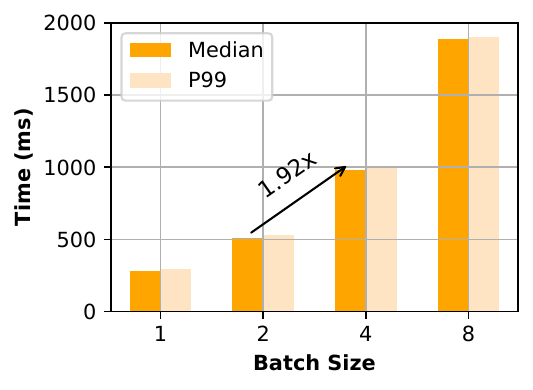}
    \caption{Image encoding time varying batch sizes.}
    \label{fig:vary-encoder-batch-mllama}
  \end{subfigure}
  \hfill
  \begin{subfigure}[b]{0.33\textwidth}
    \centering
    \includegraphics[width=\textwidth]{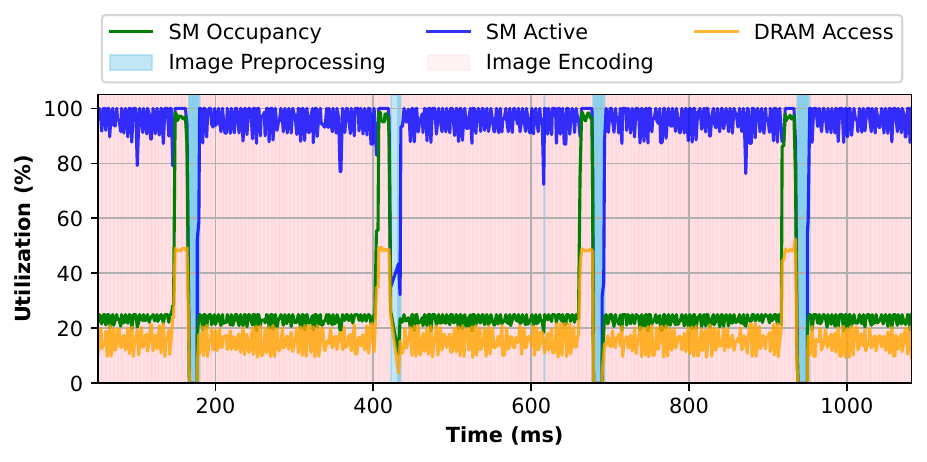}
    \caption{GPU utilization during image preprocessing and encoding.}
    \label{fig:sm-metrics}
  \end{subfigure}
  \caption{Compute characteristics of image preprocessing and encoding.
  Both stages are compute-bound.}
  \vspace{-10pt}
  \label{fig:prep-encoder-compute}
\end{figure*}

\myparagraph{Per-stage Latency Breakdown}
\Cref{fig:latency_breakdown} plots the split-up of TTFT across the three stages that comprise it; image preprocessing, image encoding, and LLM prefill.
There are three key takeaways.
First, image preprocessing, which occurs on the CPU, contributes minimally to the overall TTFT, while image encoding time contributes to a major portion in TTFT (especially for CroAttn models).
For instance, 79\% and 65\% of TTFT in Llama3.2-11B and Llama3.2-90B are from image encoding.
For DecOnly models such as InternVL-26B and NVLM-D-72B, image encoding latency accounts for 25\% and 54\% of TTFT. Second, the image encoding time depends on the encoder model size. For instance, scaling from SigLIP-400M (in LLaVA-OV-7B) to InternViT-6B (in InternVL-26B), the median image encoding time increases by 10$\times$.
Since connectors~\cite{zhu2025connector} are extremely small (\eg{}, $<0.1\%$ of total parameters in InternVL-26B), they contribute negligible latency ($<0.4\%$ of TTFT).
Finally, prefill computation is more efficient in CroAttn models because image tokens are attended to only in the CroAttn layers while the majority of LLM backends are self-attention layers, as described in \cref{sec:background:lmm}.

Taken together, these results underline why image encoding emerges as a major bottleneck in multimodal model serving. First, the number of images per request can be substantial, especially in video understanding or multi-image tasks. Second, CroAttn models shift the computational load to the image encoder by reducing image-text interaction in the LLM. This
lowers the computational load from LLM and keeps it in encoding.

\boxinsight{A major portion of the TTFT is spent on image encoding, particularly for CroAttn models, making image encoding optimization critical to meet TTFT SLOs.}
\label{insight:1}

\myparagraph{Compute Characteristics of \lmm{} Stages}
Image preprocessing on CPU and image encoding on GPU are compute-intensive processes.
\Cref{fig:vary-preprocessor-core} plots the impact of varying the number of CPU cores on preprocessing latency. Preprocessing is CPU-intensive and benefits from trivially parallelizing across cores.
Both stages exhibit linear latency scaling with batch size, saturating compute without significant throughput gains from increased batching as shown in~\Cref{fig:vary-img-preprocessor-batch,fig:vary-encoder-batch-mllama}, respectively.
\Cref{fig:sm-metrics} further plots the GPU utilization metrics for a request batch size of one during image preprocessing and image encoding.
We observe a consistent SM core activity near 100\% during image encoding, with average DRAM utilization below 30\%.
Image encoding is, therefore, typically compute-bound, resembling the language model's prefill phase~\cite{pod-attention}.
Moreover, when a request has multiple images in the input prompt (e.g., video workloads), there is typically no compute dependency between the images during image encoding; hence, image tiles can be parallelized across multiple encoders.

\boxinsight{Image preprocessing and encoding are both compute-intensive similar to LLM prefill stage.
The independence of image computations in a multimodal request enables parallelization of image preprocessing and encoding across multiple instances.}
\label{insight:2}

Compute characteristics of prefill and decode phases of the LLM backend have been well studied;
the prefill phase is typically compute-bound, while the decode phase is memory-bound~\cite{patel2024splitwise,sarathi-serve,pod-attention}.
However, \Cref{fig:latency_breakdown} shows that LLM prefill is more efficient in CroAttn models than in DecOnly models, resulting in reduced compute intensity and an interesting tradeoff we describe below.

\begin{figure}[!t]
    \centering
    \includegraphics[width=0.975\linewidth]{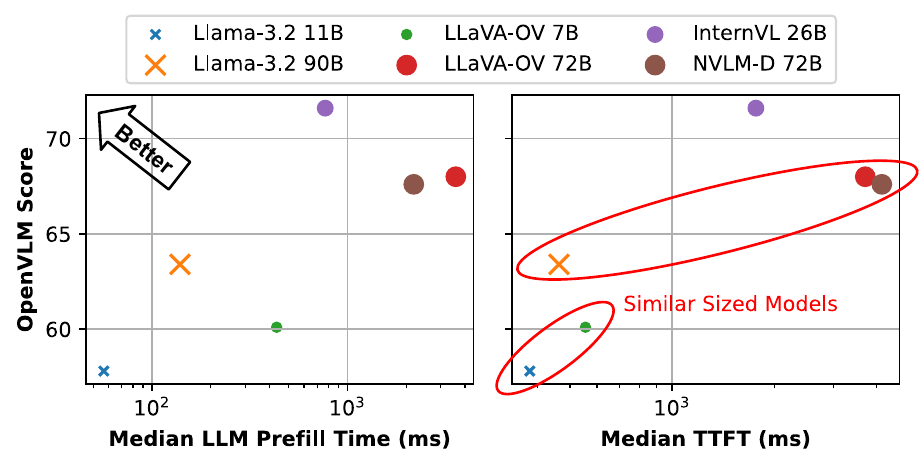}
    \caption{\lmm{} accuracy vs. prefill/TTFT efficiency.
    }
    \label{fig:scatter-plot-lmms}
\end{figure}

\myparagraph{Latency-Accuracy Profiles across \lmms{}}
\Cref{fig:scatter-plot-lmms} shows the accuracy versus prefill/TTFT efficiency for different models.
When comparing models with similar language model backend sizes across both architectures (\eg{}, Llama3.2-11B vs. LLaVA-OV-7B and Llama3.2-90B vs. LLaVA-OV-72B vs. NVLM-D-72B), we observe that CroAttn models typically have up to an order of magnitude lower LLM prefill time, leading to lower TTFT.
However, the CroAttn models usually achieve 5 points lower accuracy compared to their DecOnly counterparts on the Open VLM leaderboard~\cite{vlm-leaderboard}.
For example, Llama3.2-90B scores 63.4, while the similarly sized LLaVA-OV-72B scores 68, but with significantly higher prefill latency and TTFT than Llama3.2-90B.

\boxinsight{DecOnly models exhibit 10$\times$ worse LLM prefill latency than similar-sized CroAttn models, leading to less TTFT SLO headroom for the image encoding and thus necessitating higher scalability for image workloads.}
\label{insight:3}

\begin{figure}[!t]
    \centering
    \includegraphics[width=\linewidth]{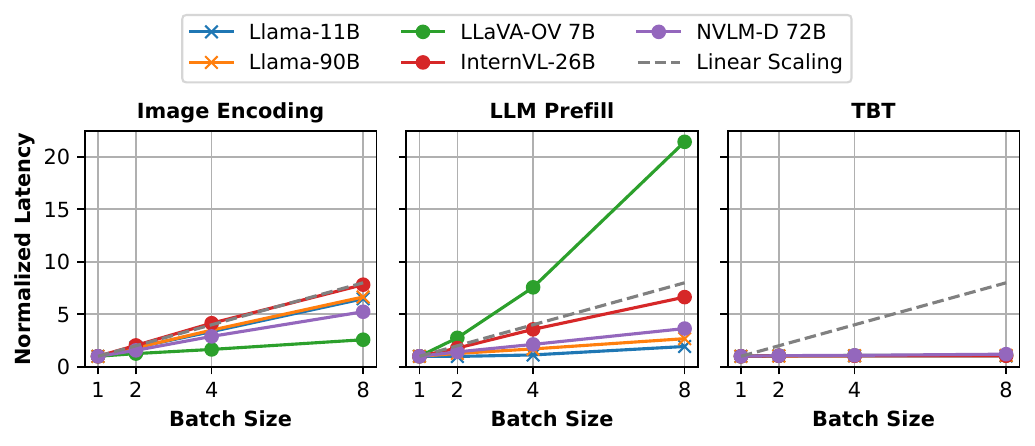}
    \vspace{-10pt}
    \caption{Median latency vs. batch size per \lmm{} stage on GPUs. Latency is normalized to that at batch size one.
    }
    \vspace{-10pt}
    \label{fig:batch-size}
\end{figure}

\myparagraph{Impact of Batching}
In today’s monolithic deployments, a single batch size is applied across all stages of the \lmm{} on the GPU, which does not strike a balance between latency and throughput given heterogeneous compute characteristics observed across different stages.
\Cref{fig:batch-size} shows how the batch size affects the median latency of each \lmm{} stage across architectures.
As the batch size increases, the latency grows at varying rates, reflecting each stage's differing sensitivity to batch size and compute intensity.

Compute-intensive stages like image encoding and LLM prefill (in DecOnly models with longer image token inputs) show limited throughput gains and rising latency beyond small batch sizes.
In contrast, the memory-bound decode stage benefits from linear throughput scaling.
Due to their low text token count, CroAttn models uniquely gain from prefill batching, diverging from traditional LLM trends where prefill saturates compute even at a batch size of one.
Notably, DecOnly model NVLM-D with fewer image tokens also exhibits certain benefits in batching.

\boxinsight{The effectiveness of batching varies for each \lmm{} component and is model-specific.
\lmm{} request batching should thus be tailored to each stage.
}
\label{insight:4}

\begin{figure}[!t]
    \centering
    \includegraphics[width=0.95\linewidth]{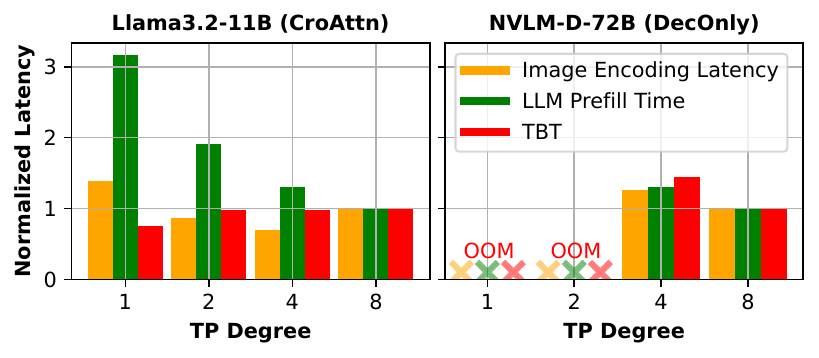}
    \caption{Impact of the tensor parallelism (TP) degree on the median latency of each stage for CroAttn-based and DecOnly \lmms{}.
    Latency is normalized to that of TP-8.
    }
    \label{fig:model-parallelism}
\end{figure}

\myparagraph{Impact of Parallelism}
Monolith deployments also limit the flexibility of model sharding within a GPU server which is typically done through tensor parallelism (TP)~\cite{vllm-tp-pp}.
\Cref{fig:model-parallelism} shows how increasing TP degrees affects latency across \lmm{} stages.
In Llama3.2-11B, the lowest LLM prefill latency occurs at TP-8, image encoding at TP-4, and TBT at TP-1.
At TP-8, encoding latency rises due to the tradeoff between compute intensity and inter-GPU communication, making it inefficient to split a 630M-sized encoder across 8 devices.

In contrast, NVLM-D-72B, with a larger 6B image encoder, sees a $1.3\times$ latency reduction when increasing TP from 4 to 8. However, this comes with diminishing returns relative to resource cost.
To balance throughput and latency, operators can deploy two TP-4 encoders for higher throughput or one TP-8 encoder for lower latency, both using eight GPUs.

\boxinsight{Treating the image encoder and LLM backend as a monolith limits parallelism flexibility and degrades performance.
Decoupling them enables independent scaling and optimized efficiency through pipelining.}
\label{insight:5}

\subsection{Production \lmm{} Trace Analysis}
\label{sec:characterization:production}
Building on the insights from open-source \lmm{} characterization, we further analyze multimodal traffic patterns at scale, leveraging production traces from one of \provider's \lmm{} inference clusters.
The traces capture a sample of multi-tenant traffic, including both text-only and image-text requests.
Our study focuses on
(1) temporal and burstiness patterns and
(2) heterogeneity of multimodal requests.
We have released these multimodal inference traces at \url{https://github.com/Azure/AzurePublicDataset}.

\begin{figure}[!t]
  \centering
  \includegraphics[width=\linewidth]{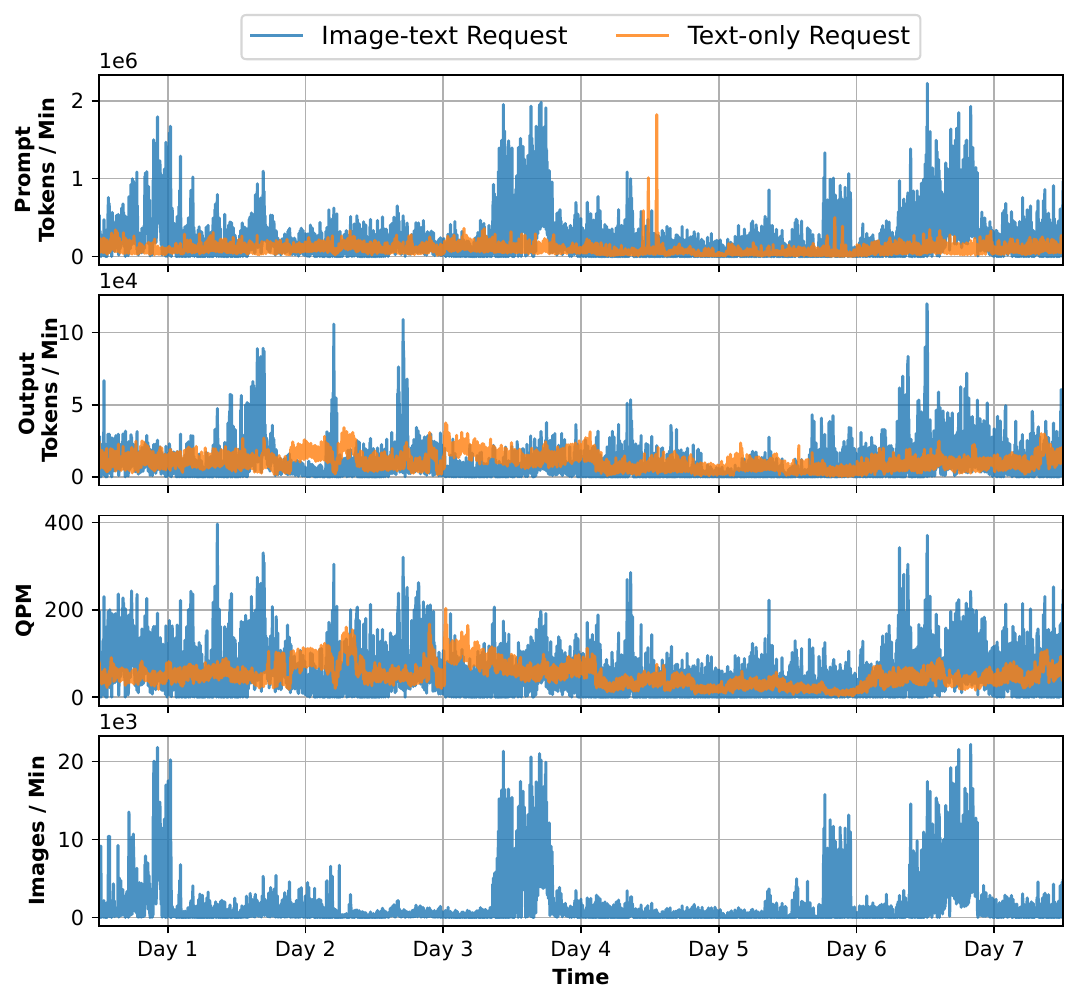}
  \caption{Aggregated prompt/out token rate, request arrival rates in queries per minute (QPM), and image rate for a production \lmm{} inference cluster in one week.
  }
  \label{fig:qpm-prod}
\end{figure}
\myparagraph{Temporal Patterns and Burstiness}
\Cref{fig:qpm-prod} shows the traffic of text-only and image-text requests separately to understand their dynamic behavior and overall impact on the system.
The traces are collected over a span of one week. 
To understand the traffic patterns, we report the timeline of prompt (input) token rate, output token rate, request arrival rate, and input image rate.
Our analysis reveals two key characteristics in production \lmm{} inference:
\begin{itemize}[nosep]
\item \myparagraphemph{Diverse Arrival Patterns}
Image-text requests show up to 5$\times$ higher prompt token rates than text-only requests.
In addition, their peak and trough occurrences are largely independent, showing minimal correlation.
\item \myparagraphemph{Image-Driven Bursts}
Image-text requests experience significant burstiness, not only due to higher request arrival rates but also increased images per request (\eg{}, video workloads). As a result, existing LLM traffic prediction methods~\cite{stojkovic2024dynamollm} (which work well for LLM workloads with diurnal patterns) have a high average prediction error rate of 79\%.
\end{itemize}

\begin{figure}[!t]
    \centering
    \begin{subfigure}[t]{0.49\linewidth}
        \centering
        \includegraphics[width=\linewidth]{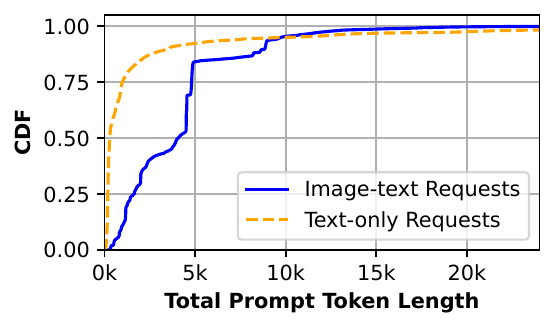}
        \caption{Total prompt length.}
        \label{fig:text-len-prod}
    \end{subfigure}
    \hfill
    \begin{subfigure}[t]{0.49\linewidth}
        \centering
        \includegraphics[width=\linewidth]{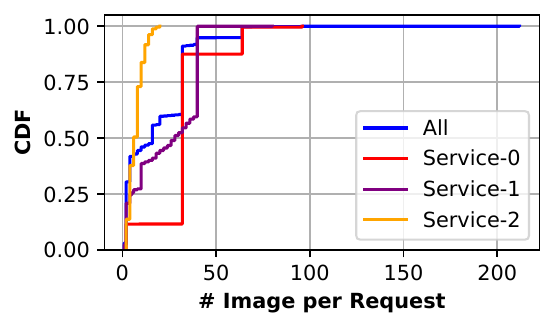}
        \caption{\#Images per request.}
        \label{fig:img-num-prod}
    \end{subfigure}%
    \caption{\lmm{} input characterization in production.
    }
    \label{fig:prod-input}
\end{figure}

\myparagraph{Request Heterogeneity} 
\Cref{fig:text-len-prod} shows that prompt lengths vary significantly across modalities.
Both image-text and text-only requests follow a heavy-tailed power-law distribution ($\alpha$=4.4 and 2.9, respectively) where a higher $\alpha$ means a heavier tail with more extreme events occurring more frequently.
In addition, image-text requests have longer median prompts due to image tokens but shorter tails than text-only requests.
\Cref{fig:img-num-prod} shows that the number of images per request also varies significantly with a heavy tail.
In addition, among the top three services issuing text-image requests, we observe high inter-service variability.
Some services (e.g., video) process 16$\times$ more images per request than others.

Comparing the image dimension distribution in our production traces with that of ShareGPT-4o image dataset~\cite{chen2024far}, we observe similar distributions, with median image width and height around 500 pixels and P95 exceeding 1000 pixels.

\boxinsight{Production \lmm{} image traffic exhibits bursty behavior independent of the traffic patterns of text requests due to the nature of different services.
Serving systems must dynamically scale resources to handle modality-specific bursts efficiently.}
\label{insight:6}

\begin{figure}[!t]
    \centering
    \includegraphics[width=\linewidth]{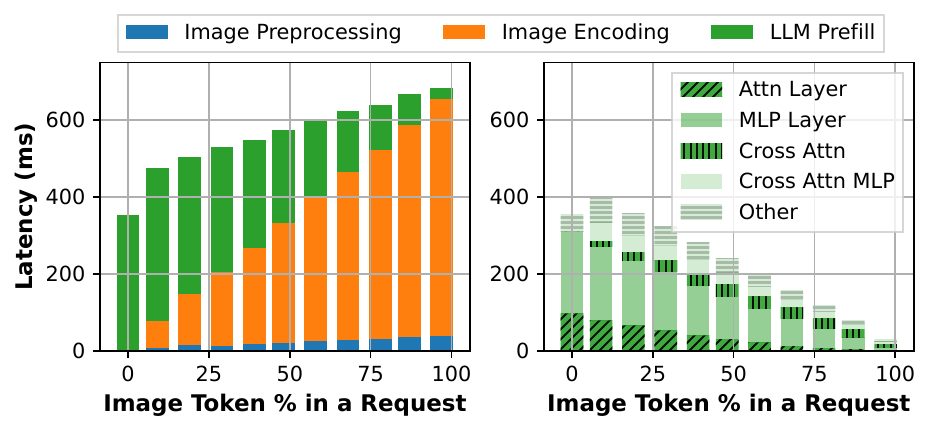}
    \caption{Llama3.2-11B (CroAttn) TTFT breakdown (left) and LLM prefill time breakdown (right) under various image-to-text token ratios in a request.
    }
    \label{fig:variation-ttft-mixed-tokens}
\end{figure}

\myparagraph{Impact of Mixed Modality}
Given \lmm{} requests' input heterogeneity, \Cref{fig:variation-ttft-mixed-tokens} shows how varying image token percentages within a single request affects TTFT and LLM prefill time in a CroAttn model Llama3.2-11B, with detailed latency breakdowns.
DecOnly models have no prefill time variation with varying token ratios as image and text tokens are treated in the same manner.
We fix the total context length of each request at 16K tokens while varying the percentage of image tokens by adjusting the number of images (0--10 images in each case with 1601 tokens per image).

TTFT increases with the percentage of image tokens in a request due to the increased image encoding computation, resulting in a 1.5$\times$ TTFT degradation when transitioning from text-only to image-only inputs.
However, this latency gain is significantly lower than DecOnly models because CroAttn models attend to image tokens only within the CroAttn layers, resulting in reduced LLM prefill time (shown in green) and partially offsetting the overhead from image encoding.
The right figure in \Cref{fig:variation-ttft-mixed-tokens} further illustrates this by breaking down the layer-wise LLM prefill time, highlighting a reduction in self-attention compute (\ie{}, ``Attn Layer'' and ``MLP Layer'') as the proportion of image tokens increases.
Although the cross-attention computation peaks at the 50\% image tokens (due to the dependency on both image and text tokens), it contributes much less than self-attention computation because there are only 4 CroAttn layers (out of 40 layers).

\boxinsight{DecOnly models maintain consistent prefill times regardless of token modality, making total token count the key factor for request routing. In contrast, CroAttn models experience reduced prefill latency as the image token percentage increases, requiring a modality-aware request routing strategy that balances both text and image token load in multimodal traffic.}
\label{insight:7}
\section{\sysname{} Design and Implementation}
Based on our insights from the characterization study of open-source \lmm{} benchmarks and production \lmm{} workloads, we propose \sysname, a novel decoupled architecture for scalable and resource-efficient \lmm{} serving.

The key idea in \sysname{} is to separate image- and text-specific inference stages into distinct instances, given the need to optimize each stage separately (\Cref{insight:1,insight:3}) and enable seamless interaction between stages.
Unlike monolithic infrastructures, \sysname{} enables independent optimization of each stage, improving resource efficiency while meeting performance SLOs.
This decoupling also enables modality-aware request serving, addressing tail latency, heterogeneous bursts, and resource contention.

\begin{figure}[!t]
    \raggedright
    \includegraphics[width=\linewidth]{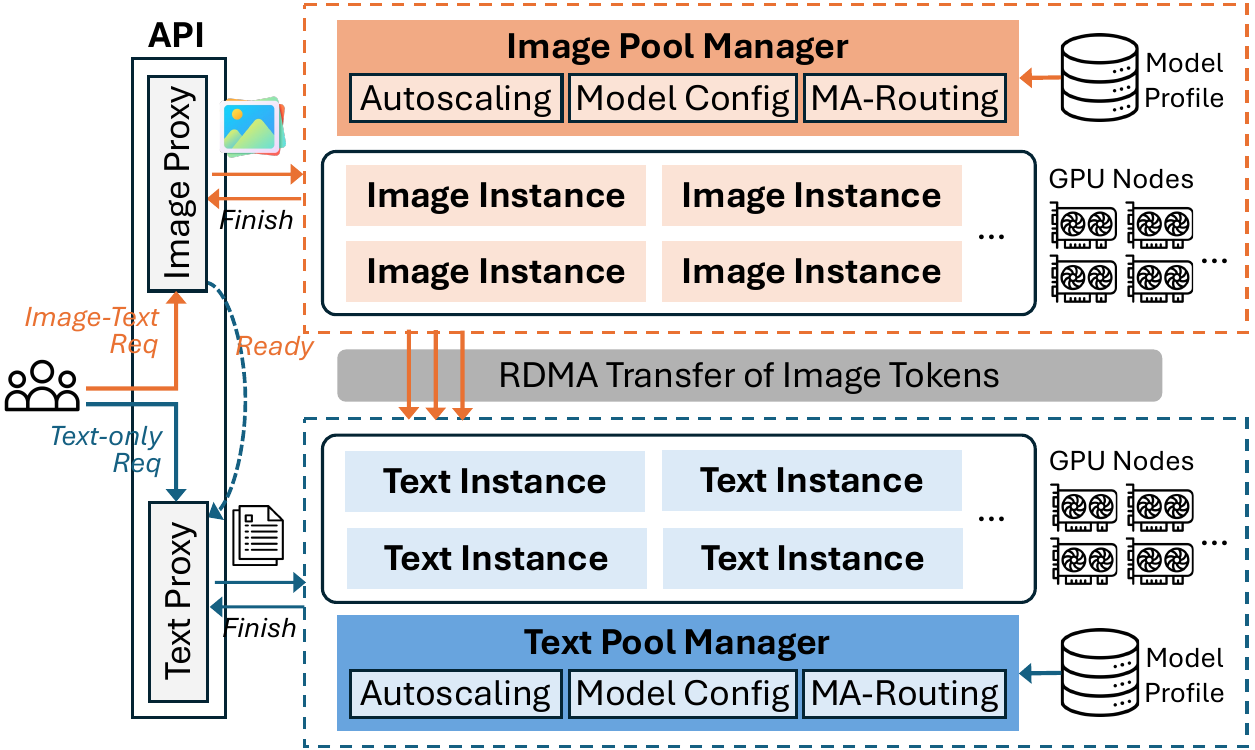}
    \caption{Overview of the \sysname{} architecture.}
    \label{fig:overview}
\end{figure}

\myparagraph{Overview}
\Cref{fig:overview} shows \sysname{}'s design.
A pool of \emph{Image Instances} handles image preprocessing and encoding of image-text requests.
The resulting image tokens are passed to a pool of \emph{Text Instances}, which performs LLM prefill and decode operations to generate outputs.
Text-only requests bypass the image components and are queued directly at the \emph{Text Instances}.
Two pools are managed by the \emph{Image and Text Pool Managers}.

\sysname{} adopts a hierarchical architecture inspired by DynamoLLM~\cite{stojkovic2024dynamollm}.
Onboarding any new \lmms{} (\eg{}, Llama3.2-11B) starts with an offline profiling phase to build model-stage profiles that capture how model configurations and load impact performance (\Cref{sec:design:profiling}).
\sysname{} uses these profiles to guide model configuration and instance scaling.
After model deployment, \sysname{} then reconfigures resources periodically to adapt to workload patterns, scaling for increased image-text requests (or vice versa) (\Cref{sec:design:resource-management}).
For each request, \sysname{} selects the optimal \lmm{} Text or Image Instances for execution (\Cref{sec:design:modality-aware-serving}).

\subsection{Offline \lmm{} Profile Generation}
\label{sec:design:profiling}
When onboarding a new \lmm{}, \sysname{} generates resource-performance profiles by characterizing the image encoder and LLM backend independently.
This profiling runs controlled inference workloads with varying model parallelisms (\eg{}, TP-2 and TP-8), batch sizes, and load (\ie{}, image tokens per second for image encoders and prompt tokens per second for LLM backends) to capture per-stage performance characteristics.
To efficiently model performance across different load conditions, \sysname{} profiles a set of representative load levels (up to the maximum throughput) and extrapolates the behavior for intermediate loads.
The resulting profiles take load, parallelism, and batch size as inputs to predict key performance metrics, including encoding latency for \emph{Image Instances} and prefill time and TBT for \emph{Text Instances}.
The \emph{Pool Managers} use these profiles to guide model-specific or architecture-specific operational decisions (\Cref{sec:design:resource-management}):
(1) pool autoscaling to meet latency SLOs without overprovisioning,
(2) model sharding that selects the optimal TP degree, and
(3) maximum batch sizing for each stage.

Since multiple \lmms{} may share the same image encoder or LLM backend, \sysname{} minimizes overhead by reusing model profiles across deployments.
These profiles are cached in cluster-local storage and synchronized via a global repository, enabling efficient sharing across clusters.

\subsection{Decoupled Resource Management}
\label{sec:design:resource-management}
\sysname{}'s decoupled approach to resource management stems from our insights on stage-specific performance disparities in batching (\Cref{insight:4}), independent scaling benefits (\Cref{insight:5}), and modality-specific traffic patterns (\Cref{insight:6}).
Specifically, \sysname{} periodically reconfigures resources (\ie{}, every five minutes to match the autoscaling overhead) to align with workload demands.

The \emph{Image Pool Manager} maintains a pool of \emph{Image Instances}, which preprocess images on CPU and encode image workloads on GPU for image-text requests.
Meanwhile, the \emph{Text Pool Manager} manages a pool of \emph{Text Instances} responsible for the prefill and decode stages of both image-text and text-only requests.
Based on model profiles (\Cref{sec:design:profiling}), each manager independently optimizes pool autoscaling, model sharding, and max batch sizing to minimize costs while meeting performance SLOs.
Since connectors are extremely small in size and contribute negligible latency ($<0.4\%$ of TTFT), dedicating separate GPUs to them would underutilize resources.
Instead, \sysname colocates connectors with the LLM backend, which accommodates cases where the connector consumes features from multiple encoders~\cite{zhu2025connector}.

Before their online operation, the initial number of Image Instances ($N_i$) is determined using the median image QPS multiplied by the median image encoding latency.
The number of Text Instances ($N_t$) is set as $N_i$ divided by the median number of images per request, based on historical \lmm{} inference traces.
If no history is available, \sysname{} initially overprovisions resources to both Text/Image Instances to ensure reliability.

\myparagraph{Token-Aware Pool Autoscaling}
The \emph{Pool Managers} dynamically scale the number of \emph{Image and Text Instances} based on real-time workload demands.
For example, a surge in image-heavy requests leads to more image preprocessors and encoders, while an increase in text requests or requests' prompt lengths triggers the \emph{Text Pool Manager} to scale LLM replicas to handle variations in prefill.

The number of replicas per stage is computed as $\lceil{\frac{ML}{MC}}\rceil$
where $ML$ is the modality-specific load (\eg{}, prompt tokens/sec for \emph{Text Instances}, image tokens/sec for \emph{Image Instances}) and $MC$ is the maximum capacity each stage can handle without violating SLOs, based on the offline \lmm{} profiling data.
Unlike traditional web service autoscaling, which reacts to request rates, \sysname{} optimizes scaling based on token throughput (tokens/s), capturing variations in both request rates and request sizes (\Cref{insight:6} and \Cref{fig:text-len-prod}).

For \emph{Image Instances}, image token counts are precomputed based on the static mapping from image dimensions to the number of tokens per image (\Cref{fig:img-token-distribution-open}).
Autoscaling of \emph{Text Instances} is based on text token load in CroAttn models but total tokens in DecOnly models due to homogeneous self-attention across modalities.
Advanced autoscaling hysteresis prevention techniques~\cite{optscaler} can be employed to avoid excessive scaling actions caused by transient workload fluctuations but are not covered in this paper.

\myparagraph{Model Sharding}
The \emph{Pool Managers} also determine instance sharding for optimal tensor parallelism (TP) for image encoders and LLM backends.
Our characterization (\Cref{sec:characterization:open-source}) shows image encoders achieve peak throughput with lower TP than LLMs.
Therefore, the model sharding degree for each instance is configured separately for maximum throughput while ensuring SLO attainment on TTFT and TBT.
By decoupling the components, \sysname{} ensures independent sharding, optimizing parallelism without unnecessary synchronization overhead.

When scaling beyond a single GPU server, \sysname{} prioritizes autoscaling over pipeline parallelism (PP)~\cite{shazeer2020megatron} to maximize throughput while avoiding communication overhead, seamlessly transitioning to batch-level optimizations as needed.

\myparagraph{Identifying Max Batch Size}
For each stage, the maximum batch size is configured to maximize throughput while meeting latency SLOs.
Batch sizing decisions are guided by the offline model-stage profiles, which predict their impact on encoding and decoding latencies.
\emph{Image Instances} may forgo batching as small max batch sizes often achieve optimal GPU utilization (\Cref{insight:3}).
In contrast, \emph{Text Instances} batch requests when beneficial, optimizing token throughput during prefill/decode based on TTFT and TBT SLOs, particularly for CroAttn \lmms{} (\Cref{insight:4}).

\subsection{Modality-Aware \lmm{} Request Serving}
\label{sec:design:modality-aware-serving}
For each incoming \lmm{} request, \sysname{} dynamically routes and schedules workloads to balance load across \emph{Image and Text Instances}.
The \emph{Pool Managers} optimize this process to minimize queueing delays and improve TTFT latency.

\myparagraph{Request Routing Across Instances}
To mitigate tail TTFT latency surges caused by modality-specific bursts and queuing delays, \sysname{} employs a modality-aware routing strategy that balances image and text workloads independently.
Traditional request-level LLM load balancing (\eg{}, round-robin, memory-based~\cite{sun2024llumnix}) overlooks the computational intensity of image encoding (\Cref{insight:2}), making them vulnerable to load imbalances during image bursts (\Cref{insight:6}), leading to high tail latencies.

Instead, \sysname{} routes requests by input modality.
Image-text requests are assigned to \emph{Image Instances} with the least image-token load.
Large requests (\ie{}, those with more images) are consequently distributed across multiple \emph{Image Instances} for parallel processing and encoding (\Cref{insight:2}), preventing degraded batching performance that would occur if all images were routed to a single instance.
This effectively enables a form of request chunking~\cite{sarathi-serve}, where images in a large request can be processed in an interleaved manner with other requests, reducing HoL blocking and improving scheduling flexibility.

To route traffic between Text Instances, text-only requests and image-text requests with completed image tokens are directed to the \emph{Text Instance} with the least total pending tokens (text+image) for DecOnly models and the least total pending text tokens for CroAttn models because of the attention mechanism difference between the two model architectures (\Cref{insight:7}).
Modality-aware routing enables parallel image encoding and dynamically adapts to image or text traffic bursts, reducing queueing delays and improving TTFT, particularly at the tail.

\myparagraph{Instance Request Scheduling}
At the instance level, \sysname{} minimizes resource contention between image-text and text-only requests with priority scheduling based on modality and prompt size.
While decoupling isolates image and text processing, contention can still arise in \emph{Text Instances}, where both request types share prefill processing.
This issue is particularly pronounced in DecOnly \lmms{}, which exhibits lower efficiency during the prefill stage (\Cref{fig:batch-size}).
Performance degradation occurs from increased batching latency for all requests, while non-batched processing introduces HoL blocking and high queueing delays at tails due to request heterogeneity (\Cref{fig:prod-input}).

To address these challenges, \sysname{} replaces traditional FIFO scheduling--which may exacerbate HoL blocking~\cite{patke2024queue,aiops2024qiu}--with an SLO-driven scheduling strategy that can prioritize shorter requests (\eg{}, text-only queries or small image-text requests with tight SLOs) to maintain low latency.

\emph{Pool Managers} continuously monitor SLO attainment and trigger pool autoscaling when the rate falls below a predefined threshold (default 0.99 with a sensitivity study in \Cref{sec:eval:sensitivity}), ensuring adaptive resource allocation under dynamic workloads (especially in cases of unpredictable traffic).
\sysname{} can work with state-of-the-art batch scheduling techniques~\cite{sarathi-serve,patel2024splitwise} to optimize TBT during the decode stage, which we leave to future work as we do not observe TBT degradation in \lmm{} characterization (\Cref{insight:4}).

\myparagraph{Image Token Transfer}
Once image encoding is complete, \sysname{} transfers image tokens from \emph{Image Instances} to \emph{Text Instances} via a pull-based RDMA mechanism.
Push-based approaches immediately transmit image tokens as they are generated, requiring premature decisions about which \emph{Text Instance} should receive them (and all subsequent tokens).
This design increases synchronization overhead and risks suboptimal routing, as the system operates with incomplete information about token counts and runtime load.

In contrast, our pull-based design defers transfer until all image tokens for a request are ready.
This many-to-one aggregation enables the scheduler to select the target \emph{Text Instance} with full information, considering factors such as queue size, prefix caching, and payloads.
Once the routing decision is made, the request carries RDMA addresses of the producing \emph{Image Instances}, from which the chosen \emph{Text Instance} pulls the image tokens.


\sysname{} colocates \emph{Image and Text Instances} in the same server when each \emph{Text Instance} is not taking up all GPUs on a server to avoid image transfer overhead and unallocated idle GPUs.
For example, it may place one TP-4 Text Instance and two TP-2 Image Instances within the same 8-GPU server.
Unlike monolithic deployments, colocated instances remain independently configurable and can serve corresponding stages of different requests independently.

\subsection{Implementation}
\label{sec:design:implementation}

We implement \sysname{} using 5,000 lines of Python code.
We base the \emph{Text Instance} on vLLM~\cite{vllm} (v0.7.2), a state-of-the-art generative model inference platform, and build the \emph{Image Instance} on HuggingFace Transformers~\cite{hf-transformers}.
The modular architecture of \sysname{} enables easy integration with other serving engines (\eg{}, TensorRT~\cite{tensor-rt} and DeepSpeed~\cite{deepspeed}).
We use \texttt{numactl} to restrict CPU and memory usage of image preprocessing to a single NUMA node, which reduces memory access latency and performance variation.
To ensure efficient GPU-to-GPU memory transfer of image tokens, we use PyTorch's distributed communication with the NCCL backend and GPU Direct RDMA.

We implement the \emph{Image and Text Pool Managers} as lightweight gRPC servers (hosted on dedicated VMs) with low memory and compute requirements, drawing inspiration from DynamoLLM~\cite{stojkovic2024dynamollm}.
For failure detection and recovery, the \emph{Pool Managers} use a simple heartbeat-based membership management protocol~\cite{gupta2001scalable}.
However, \sysname{} can be easily extended to adopt more robust leader election (\eg{}, Raft~\cite{raft}) and fault-tolerance algorithms~\cite{pb}.
\section{Evaluation}
\label{sec:eval}

\subsection{Experimental Setup}
\label{sec:eval:setup}

\myparagraph{Models and Workloads}
We use two representative multimodal models, Llama3.2-11B and InternVL-26B, for CroAttn and DecOnly \lmms{}, respectively.
To ensure realistic workload distribution, we adopt the inter-arrival timestamps of requests and the number of images associated with each request (ranges from 0 to 16) from the production LMM inference trace (\Cref{sec:characterization:production}) and reuse the \lmm{} text-image dataset from \Cref{sec:characterization:open-source}.

\myparagraph{Hardware}
We evaluate \sysname{} on a cluster with 16 DGX-A100 servers~\cite{a100azure} (128 GPUs).
Each server has the same configuration as the server used in our characterization study (\Cref{sec:characterization:open-source}).
The GPUs within a server are connected with NVLINK 3.0 while cross-server connection is via InfiniBand.

\myparagraph{Baselines and Systems}
We compare \sysname{} against the state-of-the-art generative model inference serving system, vLLM~\cite{vllm}, which supports \lmm{} inference as a monolithic setup.
We also evaluate \sysname{} with a few variants of \sysname{} implemented on top of vLLM:
(1) vLLM with decoupled Image/Text Instances (\ie{}, \sysname{}-Decoup),
(2) \sysname{}-Decoup plus modality-aware scheduling (\ie{}, \sysname{}-Sched), and
(3) \sysname{}-Sched plus modality-aware routing (\ie{}, \sysname{}), for ablation study.

\myparagraph{SLO Definition}
We define the SLO metrics for \lmm{} inference based on the TTFT/TBT during the isolated run of a single text-only request and text-image request (with one image) on the monolith baseline setup.
Then, we scale the SLO metrics with a constant factor (\ie{}, SLO factor) to evaluate how \sysname{} performs under tight/relaxed SLOs (\Cref{sec:eval:sensitivity}).
The SLOs are defined on P99 tail latency across requests over time.

\subsection{End-to-end Performance}
\label{sec:eval:e2e}

\begin{figure}[!t]
  \centering
  \includegraphics[width=\linewidth]{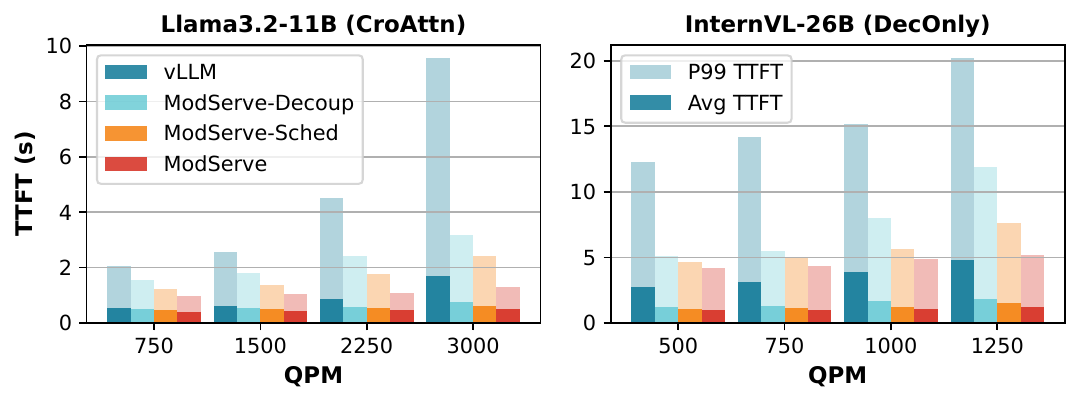}
  \caption{TTFT comparison with fixed 16 servers (128 GPUs) without autoscaling.}
  \label{fig:eval:static-latency}
\end{figure}

\begin{figure}[!t]
    \centering
    \includegraphics[width=\linewidth]{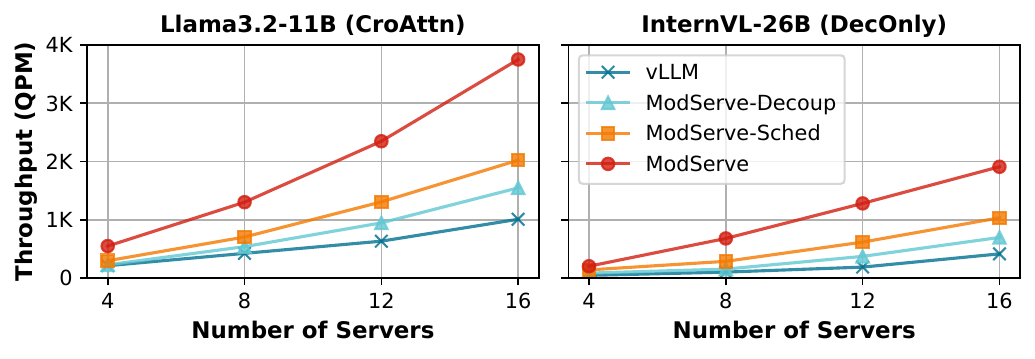}
    \caption{Maximum load meeting SLO.}
    \vspace{-10pt}
    \label{fig:eval:static-capacity}
\end{figure}

\myparagraph{Static Resource Allocation}
We begin by evaluating \sysname{} under a static resource allocation setup, where a fixed number of servers remain active at all times without autoscaling.
This setup isolates the benefits of decoupling, modality-aware request scheduling, and routing from pool autoscaling (which we explore independently).
\Cref{fig:eval:static-latency} shows the average and tail (P99) TTFT achieved by \sysname{} and the baselines when serving different input loads over fixed resources (16 servers with 128 GPUs in total).
In this setup, vLLM (monolith) deploys 32 instances (each with TP-4) while the other approaches (decoupled) deploy 20 Text Instances (TP-4) and 48 Image Instances (TP-1).

Compared to vLLM, statically decoupling (\sysname{}-Decoup) improves the average and P99 TTFT by 27\% and 42\% (for Llama3.2), 46\% and 47\% (for InternVL).
This is because monolithic deployments process all modalities on shared GPU resources, leading to contention and inefficient utilization under imbalanced modality traffic. 
In addition, \sysname{}-Decoup with the same number of GPUs can deploy 16 extra Image Instances and enables image encoding parallelization that reduces TTFT significantly compared to the monolithic deployment on vLLM.

\sysname{} shows a more pronounced TTFT improvement over the monolith baseline when serving InternVL.
This is because the monolith deployment faces resource contention with DecOnly models due to their high prefill latency (\Cref{insight:3}), which contends with image encoding.
Additionally, InternVL's image encoder has higher batching performance degradation (\Cref{insight:4}) and thus benefits more from parallelization.
Adding modality-aware request scheduling (\sysname{}-Sched) further reduces the average and P99 TTFT by 12\% and 25\%, modality-aware routing (\sysname{}) reduces the average and P99 TTFT by 14\% and 32\%, as it reduces HoL blocking and mitigates tail latency spikes.

Overall, \sysname{} achieves the lowest TTFT across all load levels, demonstrating the effectiveness of modular inference pipelines.
We observe similar TBT performance in all approaches due to its compute insensitivity (as indicated by \Cref{fig:batch-size}).
\Cref{fig:eval:static-capacity} further evaluates the maximum throughput under the TTFT and TBT SLO when varying the static resource allocation from 4 to 16 servers.
\sysname{} achieves a 3.3$\times$ and 5.5$\times$ throughput improvement over vLLM (monolith) for Llama3.2 and InternVL, respectively, which confirms that DecOnly models benefit more from decoupling. 

\begin{figure}[!t]
  \centering
  \includegraphics[width=\linewidth]{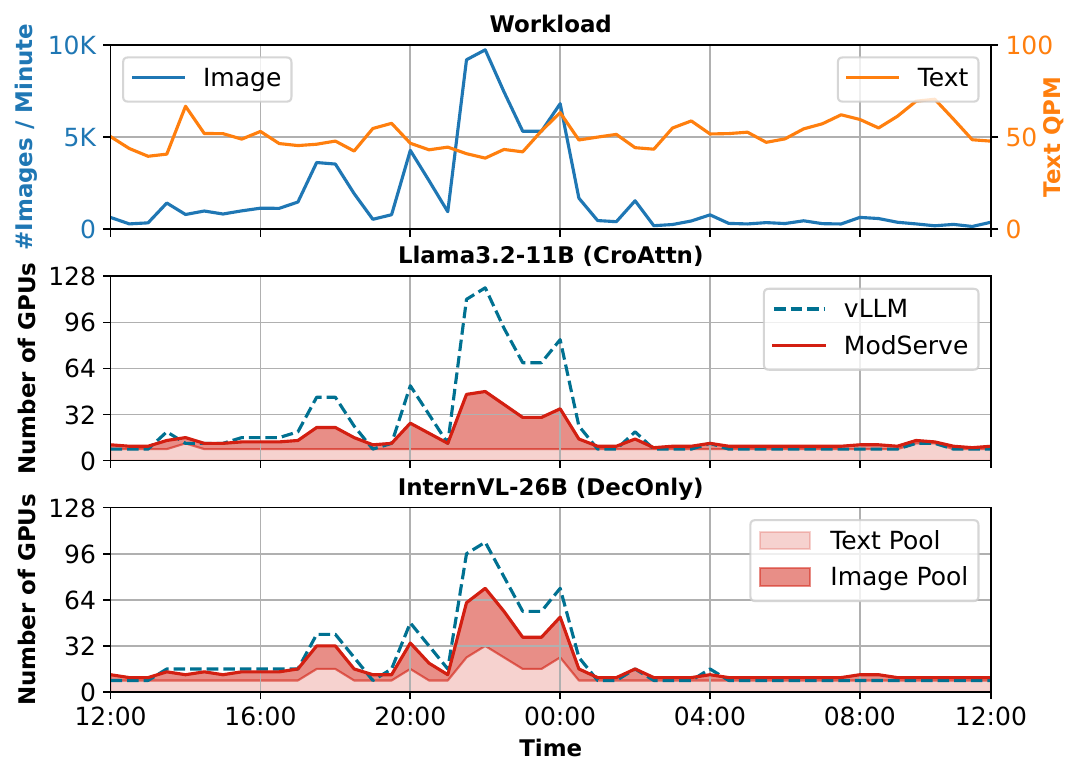}
  \caption{GPU allocation with autoscaling (up to 16 servers) during a one-day interval on the production traces.}
  \label{fig:eval:autoscaling}
\end{figure}

\myparagraph{Resource Allocation with Autoscaling}
We now assess how \sysname{} and vLLM (monolith) baseline handle image-driven bursts seen in the production trace (\Cref{fig:qpm-prod}).
Fundamentally, to serve traffic bursts, a system needs to scale up the resources to meet the workload demand while scaling down to avoid overprovisioning.
Therefore, we enable autoscaling in both \sysname{} and vLLM and evaluate them on a one-day interval of the production trace that contains an image-driven burst.
For a fair comparison, both \sysname{} and vLLM (monolith) use similar SLO-driven autoscaling heuristics based on offline model profiling (\Cref{sec:design:resource-management}).

\Cref{fig:eval:autoscaling} compares the number of GPUs used by \sysname{} and vLLM (monolith) to serve the image-driven burst in the production trace. \sysname{} takes 41.3\% and 25\% fewer GPUs compared to vLLM to serve Llama-3.2 (CroAttn) and InternVL (DecOnly) models respectively while meeting the tail latency SLOs.
\sysname{}'s cost reduction is higher for Llama-3.2 (CroAttn) model because the increase in image tokens caused by image-driven traffic bursts does not overwhelm the LLM backends in CroAttn models as observed in its latency profile (\Cref{fig:variation-ttft-mixed-tokens}).
However, in InternVL (DecOnly), the LLM backend's latency increases with the increase in image tokens due to homogeneous self-attention.
Therefore, to meet SLOs, \sysname{} scales up the number of Text Instances for InternVL more than for Llama-3.2 during image-driven bursts (light pink in \Cref{fig:eval:autoscaling}).
Overall, \sysname{}'s stage-aware autoscaling prevents unnecessarily scaling up LLM backends (done by vLLM due to monolith deployment) during image-driven bursts and prevents resource over-provisioning.

\subsection{Sensitivity Study}
\label{sec:eval:sensitivity}

\begin{figure}[!t]
    \centering
    \includegraphics[width=\linewidth]{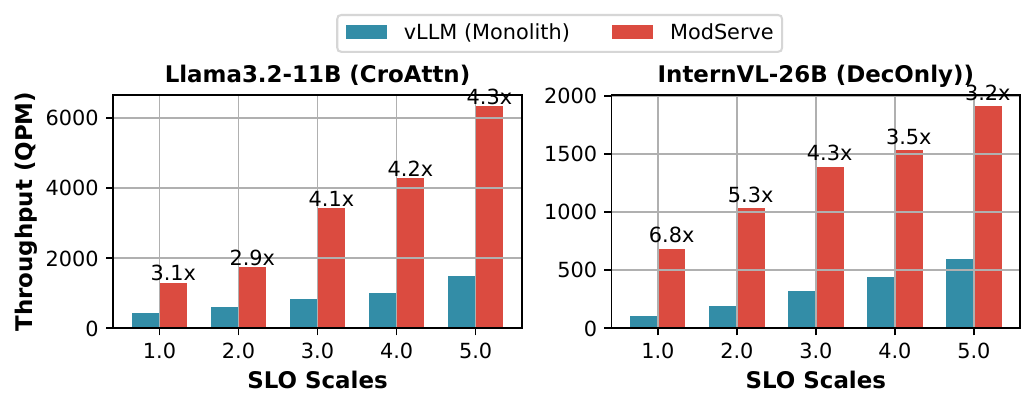}
    \caption{Throughput impact varying the SLO scale.}
    \vspace{-10pt}
    \label{fig:eval:sensitivity-slo-scale}
\end{figure}

\myparagraph{Impact of SLO Scale}
\Cref{fig:eval:sensitivity-slo-scale} shows the maximum throughput \sysname{} can achieve when changing the SLO scale (higher values refer to more relaxed SLOs).
As the SLO scale increases, \sysname{} consistently outperforms the vLLM, achieving up to 4.3$\times$ higher throughput for Llama-3.2 and 6.8$\times$ for InternVL.
This trend highlights that \sysname{} better utilizes resources under the same latency requirements.

\begin{figure}[!t]
  \centering
  \begin{subfigure}[b]{\linewidth}
    \centering
    \includegraphics[width=\linewidth]{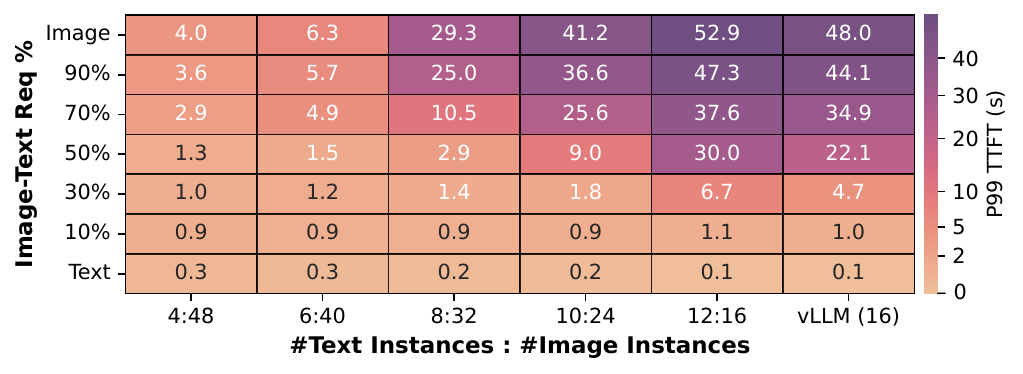}
    \caption{Llama3.2-11B CroAttn \lmm{}.}
    \label{fig:eval:sensitivity-ratio-llama}
  \end{subfigure}%
  \hfill
  \begin{subfigure}[b]{\linewidth}
    \centering
    \includegraphics[width=\linewidth]{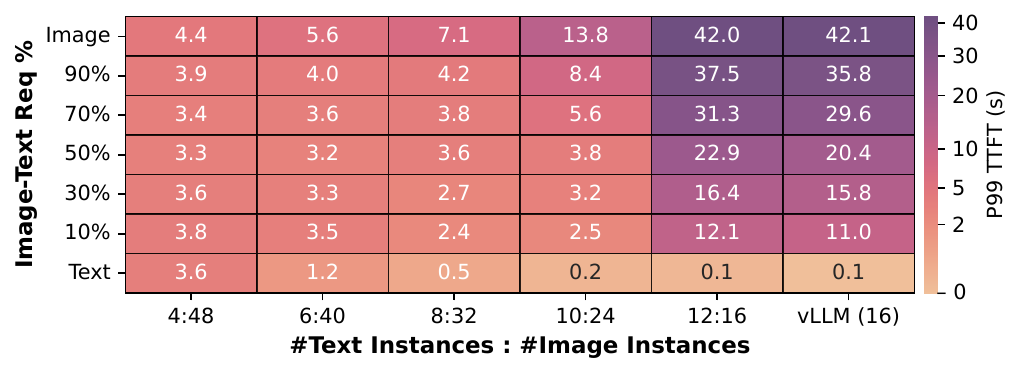}
    \caption{InternVL-26B DecOnly \lmm{}.}
    \label{fig:eval:sensitivity-ratio-internvl}
  \end{subfigure}%
  \caption{Impact of image request percentage ($Y$-axis) and instance allocation ($X$-axis), \ie{}, \#Text Instances (TP4) : \#Image Instances (TP1) on 8 servers (64 GPUs).}
  \vspace{-10pt}
  \label{fig:eval:sensitivity-ratio}
\end{figure}

\myparagraph{Impact of Image-to-Text Instance Ratio}
\Cref{fig:eval:sensitivity-ratio} shows the effect of varying the ratio of Image and Text Instances on 64 GPUs (8 servers) along the $X$-axis, in comparison to vLLM monolith with 16 instances.
For instance, ``4:48'' denotes a configuration with 4 Text Instances (TP-4) and 48 Image Instances (TP-1).
As the ratio of Text Instances increases, we observe that \sysname{} consistently achieves superior TTFT performance compared to vLLM (monolith) until the ratio reaches 10:24.
However, at 12:16, the decoupled configuration contains the same number of image encoders but 4 fewer LLM backends, resulting in inferior performance.
Moreover, reducing image encoders below the monolith baseline contradicts the core goal of decoupling to scale up/out the image encoders independently for multimodal processing.

\myparagraph{Impact of Image:Text Request Ratio}
\Cref{fig:eval:sensitivity-ratio} also shows the impact of varying image-text request percentages in the workload ($Y$-axis).
As this percentage increases from 10\% to 90\% (more image-heavy), TTFT for Llama-3.2 (CroAttn) increases.
InternVL (DecOnly) follows a similar trend, except at lower \emph{Text Instance} ratios (\eg{}, 4:48), where P99 TTFT decreases from 3.8 to 3.3 seconds due to reduced text load.
This stems from DecOnly models' poor prefill efficiency.
For the same reason, at low image-text request percentages (\eg{}, 10\%), InternVL sees a lower P99 TTFT as more \emph{Text Instances} help distribute the text-heavy load.

On the other hand, across all image-text request percentages, increasing the number of \emph{Text Instances} raises P99 TTFT in Llama3.2 due to a reduced number of \emph{Image Instances}, leading to longer image encoding times.
However, regardless of distribution, \sysname{} outperforms the monolith baseline (by up to 18.4$\times$ for Llama3.2 and 9.2$\times$ for InternVL) when Image:Text Instance ratio exceeds 2.4, demonstrating its efficiency handling multimodal workloads.



\begin{figure}[!t]
  \centering
  \includegraphics[width=\linewidth]{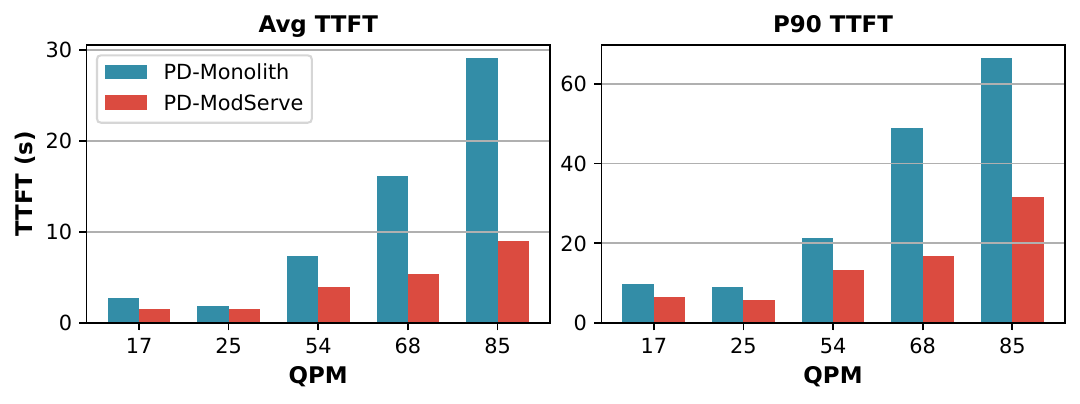}
   \caption{TTFT improvement with \sysname{} from a prefill-decode disaggregated setup for InternVL-26B (DecOnly).}
  \vspace{-10pt}
 \label{fig:eval:pd-disagg}
\end{figure}

\myparagraph{Model Architectures}
Our evaluation on open-source \lmms{} includes models up to 90B parameters, while production deployments may involve even larger model sizes affecting image encoding ratios in TTFT, which we defer to future work.
We focus on visual LMMs but audio-based multimodal models~\cite{attt} share similar architectures and parameter scales with vision multimodal models.
We also note that hybrid multimodal architectures have been proposed~\cite{nvlm}, though no open-source hybrid models are currently available.

\subsection{Prefill-Decode Disaggregation Support}
\label{sec:eval:pd-disaggregation}
\sysname is complementary to existing techniques for LLM backend optimization, including prefill/decode (PD) disaggregation~\cite{patel2024splitwise,zhong2024distserve}, which splits LLM inference into two execution phases: prefill and decode (token-by-token generation).
Our design fully supports PD disaggregation, which leads to a full EPD disaggregation.

To demonstrate this, we compare two deployment configurations under varying load, both incorporate PD disaggregation, deploy the InternVL-26B model, and use the same number of decode instances to match TBT latency (orthogonal to \sysname's contributions).
The main difference between the two configurations comes in the LLM prefill instances: 
(1) {PD-Monolith}: 4 prefill instances are deployed, where each instance is distributed across 8 GPUs. Each prefill instance also hosts an image encoder for image preprocessing and encoding.
(2) {PD-\sysname}: 3 prefill instances are deployed, each across 8 GPUs. Image encoders are fully decoupled from the LLM backends and run as 8 independent processes on the remaining GPU server.
Both configurations use a total of 32 GPUs for image encoding and LLM prefill combined.

This setup allows us to isolate the benefits of stage-level decoupling in \sysname from PD disaggregation.
\Cref{fig:eval:pd-disagg} demonstrates that \sysname (blue) provides additional TTFT reduction (up to 2.8\texttimes{} in average TTFT and 3.2\texttimes{} in P90 TTFT for InternVL-26B) beyond what PD disaggregation alone can offer (red).
The TTFT improvement (for both mean and P90) becomes more pronounced when load increases as \sysname{} reduces resource contention between the image encoding and LLM prefill stages and leverages encoder parallelization to reduce encoding latency (\cref{insight:2}).

\subsection{Token Transfer Overhead}
\begin{figure}[!t]
    \centering
    \includegraphics[width=0.8\linewidth]{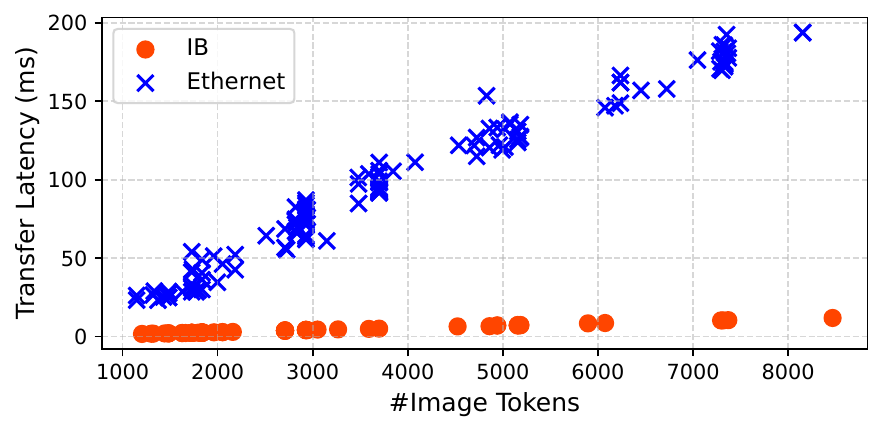}
    \caption{Image token transfer latency across token sizes}
    \label{fig:transfer}
\end{figure}

\Cref{fig:transfer} shows the image token transfer overhead for varying-sized image embeddings, comparing different communication media of using Infiniband and Ethernet.
With RDMA on Infiniband, we observed the P99 transfer latency of image tokens per image request is 5 ms, which corresponds to <0.5\% and <0.3\% TTFT for CroAttn and DecOnly models, respectively.
TCP over Ethernet incurs significantly higher overheads, with a P50 of 100 ms and a P99 of 180 ms.
\sysname{} supports both communication media. When evaluated over TCP, \sysname{} achieves a 35\% TTFT reduction at high load for InternVL-26B and an 8.4\% reduction at low load compared to the monolithic baselines (with Infiniband, the reduction is 46\% and 13\%, respectively, as mentioned in \Cref{sec:eval:e2e}).

\section{Related Work}

\myparagraph{\lmm Characterization}
Lee \etal{}~\cite{lee2024characterizing} provides a comprehensive characterization of multimodal \textit{generation} models at Meta, while we focus on \lmms{} with multimodal input (\eg{}, visual understanding models).
Hou \etal{}~\cite{hou2022characterizing} focus on traditional multimodal models employing small-scale convolutional neural networks.
In contrast, our work presents a detailed analysis of multimodal input workloads on both open-source \lmm models and production traces, highlighting their unique execution and workload patterns.

\myparagraph{\lmm Serving Optimization}
Recent research has introduced several techniques to optimize \lmm serving by addressing key inefficiencies in inference computation and memory usage.
Inf-MLLM~\cite{ning2024infmllm} employs token caching strategies and attention bias to maintain performance with long contexts while reducing KV cache memory consumption.
Elastic Cache~\cite{liu2025efficient} utilizes an importance-driven cache merging strategy to prune KV caches efficiently during inference.
Dynamic-LLaVA~\cite{huang2024dynamic}, VTW~\cite{lin2024boosting}, and QueCC~\cite{li2024inference} present various vision token sparsification and compression techniques to dynamically reduce redundancy in vision tokens, especially for video workloads.
These optimizations primarily operate at the model level, trading off computational overhead with output quality (\ie{}, accuracy).
They are orthogonal to our proposed system-level design for inference efficiency that does not impact model accuracy, which can further benefit from such model-level advancements, \eg{}, faster image encoding with subsampling~\cite{lei2021less}.

To optimize \lmm{} inference, concurrent works adopt a similar stage decoupling idea (\eg{}, EPD~\cite{epd} and HydraInfer~\cite{dong2025hydrainfer}) and parallel encoding (\eg{}, IRP~\cite{epd}).
In contrast, our work extends beyond stage decoupling by incorporating stage-aware model configuration, modality-aware routing, and autoscaling, rooted in insights from a comprehensive systems analysis of production \lmm{} inference workloads.
In addition, our characterization and evaluation take a closer look at two representative \lmm{} architectures, rather than being limited to decoder-only models.

\myparagraph{Text-Centric LLM Serving}
Recent studies have delved into disaggregating LLM prefill and decode phases for text-only LLM serving.
Examples include Splitwise~\cite{patel2024splitwise}, DistServe~\cite{zhong2024distserve}, Mooncake~\cite{qin2024mooncake}, and MemServe~\cite{hu2024memserve}.
Other optimizations for LLM serving include key-value cache management~\cite{vllm}, continuous batching~\cite{orca}, request scheduling~\cite{aiops2024qiu,patke2024queue,sun2024llumnix,sarathi-serve,agrawal2024medha}, and energy optimization~\cite{stojkovic2025tapas,stojkovic2024dynamollm,qiu2024muserve}.
While these optimizations can be applied in \sysname{} to enhance LLM backend prefill and decode efficiency, our work focuses on the unique characteristics of multimodal models.
\section{Conclusion}

We present the first comprehensive systems analysis of \lmms{} on both open-source models and production \lmm{} inference traces.
Our insights lead to the design of \sysname{}, a scalable and resource-efficient \lmm{}-serving framework that decouples inference stages for dynamic reconfiguration, adaptive scaling, and modality-aware scheduling.
Evaluations show that \sysname{} achieves 25--41\% cost savings compared to the state-of-the-art while efficiently serving production-scale \lmm{} inference workloads.

\section*{Acknowledgments}
This project was partially supported in part by U.S. NSF grants NSF-2421782, NSF-2350425, NSF-2319988, NSF-2206522, Microsoft Research Faculty Fellowship 8300751, and the Commonwealth Cyber Initiative (CCI), an investment in the advancement of cyber research, innovation, and workforce development. For more information about CCI, visit \href{cyberinitiative.org}{cyberinitiative.org}.

\bibliographystyle{ACM-Reference-Format}
\bibliography{references}

\end{document}